\documentclass[aps,prl,preprint,superscriptaddress,showpacs,showkeys]{revtex4-1}
\usepackage{amsmath}
\usepackage{amsfonts}
\usepackage{amssymb}
\usepackage{bm}

\begin{document}

\title{On the theory of the Schr\"{o}dinger equation with the full set \\
of relativistic corrections}

\author{A. A. Eremko}
\email[]{eremko@bitp.kiev.ua}
\thanks{}
\affiliation{Bogolyubov Institute for Theoretical Physics, Metrolohichna str., 14-b, 
 Kyiv,  Ukraine, 03143}
\noaffiliation
\author{L. S. Brizhik}
\email[]{brizhik@bitp.kiev.ua}
\homepage[]{https://sites.google.com/view/larissabrizhik}
\thanks{}
\affiliation{Bogolyubov Institute for Theoretical Physics, Metrolohichna str., 14-b, Kyiv,  Ukraine, 03143}
\author{V. M. Loktev}
\email[]{vloktev@bitp.kiev.ua}
\thanks{}
\affiliation{Bogolyubov Institute for Theoretical Physics, Metrolohichna str., 14-b, 
 Kyiv,  Ukraine, 03143}
\affiliation{National Technical University of Ukraine "KPI", av. Peremohy 37,
 Kyiv, Ukraine, 03056
}

\begin{abstract}

All relativistic corrections to the Scr{\"o}dinger equation which determine the interlink between spin and orbit of moving particles, are directly calculated from the Dirac equation using the spin invariant operators. It is shown that among the second order corrections there are not only the well-known Darwin and Thomas terms, but also the new ones. Only with the account of the latter corrections the energies found with the obtained spin-orbit interaction operator, coincide with the energies of the Dirac equation exact solution. The problem of electron spectrum in the quantum well type structures is studied in details and the physical reasons for the appearance of spin-orbit interaction operators in the Dresselhaus or Rashba form, are analyzed. 

\end{abstract}


\maketitle

\nopagebreak

%\section{Introduction } \label{Intr}
\section{1. Introduction} 

Study of the spin-orbit interaction (SOI) is one of the main streams of modern solid state physics \cite{Fabian} which leads to important practical applications, such as an effective control tool of spin-polarized carrier states in spintronics devices. In particular, spin spliting  arising from Rashba SOI \cite{Rashba} allows manipulating spin in semiconducting heterostructures by electric field \cite{Fabian,Winkler}. On the other hand, SOI is also the source for some interesting physical phenomena such as spin current and Hall effects. At last, SOI determines the peculiarities of a new class of condensed systems, so called topological dielectrics. According to Rashba's recent remark  \cite{Rashba16}, SOI as the notion and physical reality, "goes global",  deeply penetrating into many areas of the fundamental science or technical applications, providing new phenomena, on which future technologies will be based.  

Functioning of spintronics devices at ambient conditions (atmospheric pressure and room temperature) require strong enough spin splitting, and, hence, large SOI \cite{Winkler,Rashba16,Ishizaka,Crepaldi,Eremeev}. Among such materials there are two-dimensional (2D) or quasi-2D systems, such as layered structures and heterostructures,  crystal surfaces, interfaces, thin films (up to monomolecular or monoatomic width). The interest to low-dimensional electron phenomena as a whole, has  started growing from 1970-s, but during last 5-7 years it has increased dramatically \cite{2Dmat16} because of their potential perspectives in using in the devices of the future generation. 
 
In systems with broken inversion symmetry, Rashba splitting of 2D electron and hole bands takes place in the result of the SOI constant finite value. With decreasing system dimensionality number of carrier spatial degrees of freedom also reduces. Charge particle motion in one or two directions becomes finite and SOI proves to be one of the factors which determines the mobile carriers states. 
 
It is generally adopted that the problem of the relation of particle propagation and its spin state, described by SOI, has been solved. This includes formal derivation of SOI and understanding of physical reasons determining its existence.  Starting with the pioneer papers of Dresselhaus \cite{Dresselhaus} and Rashba \cite{Rashba1959}, account of SOI in electron structure of crystals is based on the solutions of the Schr{\"o}dinger-Pauli equation  $ \mathrm{H} \psi = E \psi $ for two-component electron wave function  (spinor) $ \psi $ with the Hamiltonian  
\begin{equation}
\label{H_nrl}
\mathrm{H} = \frac{\mathbf{p}^{2}}{2m} + V(\mathbf{r}) + \hbar ^{-1} \lambda_{SO}^{(0)}  \left( \left[\bm{\nabla}V\times \mathbf{p}\right] \bm{\hat{\sigma}} \right) , \quad  \lambda_{SO}^{(0)} = \frac{\hbar ^2}{4m^{2}c^{2}} ,
\end{equation} 
which defines their eigen energies $ E $. Here $ V(\mathbf{r}) $ is the potential, in which an electron moves,  $ \hat{\mathbf{p}} = -i\hbar \bm{\nabla} $ is its momentum operator, and  $ \bm{\hat{\sigma}} $ is the spin operator (operator of the intrinsic momentum  $ (\hbar /2) \bm{\hat{\sigma}} $), whose components are represented by Pauli matrices $ \hat{\sigma}_{j} $ ($ j = x, y, z $). The last term in the Hamiltonian (\ref{H_nrl}) is known as Thomas correction and is usually called SOI operator (Darwin SOI term  is omitted in (\ref{H_nrl}) as comparatively small). 

As it is generally known, the fundamental basis for  studying electron states is Dirac theory from  which in a natural way the existence of an electron spin $\hbar /2$ and Fermi-statistics for electrons follow. Electron spectra, calculated within this theory, practically coincide with their observable values.  Expanding Dirac equation (DE) with respect to the degrees of the ratio   $ p/mc $, where  $ p $ is characteristic momentum,  $ m $ is mass and $ c $ is the light speed, one can calculate relativistic corrections to the non-relativistic Schr{\"o}dinger equation (SchE). In other words, SchE with the Hamiltonian  (\ref{H_nrl}) is the limit of DE when particle's rest energy,  $ mc^{2} $, significantly exceeds all other energy scales. In this sense, SOI operator can be considered as one of the relativistic corrections of the order  $ 1/c^{2} $ to non-relativistic Hamiltonian \cite{Bete,RelQuant,Davydov,Messia}.

It is worth to recall that 2D electrons can be modeled as the states determined by the quantum well (QW) formed by a layer of the heterostructure or by a surface of the interface, with the potential changing in one direction only, namely in the perpendicular to the plane, $z$-direction. Free motion of carriers in $ xy$-plane is characterized by the 2D wave vector $ \mathbf{k}_{\perp} = k_{x} \mathbf{e}_{x} + k_{y} \mathbf{e}_{y} $ ($ \mathbf{e}_{j} $ are corresponding unit vectors), and 1D SchE determines discrete eigen states in the QW in which each energy level creates 2D electron band $ E_{n}\left( \mathbf{k}_{\perp} \right) $. Taking into account that the equality $ \bm{\nabla}V = \mathcal{E} \mathbf{e}_{z} $ is valid for asymmetric QW with the asymmetry arising from the electric field $ \mathcal{E} $ that is perpendicular to the QW plane, one can derive a model with the operator $ H_{R} = \lambda_{BR} \left[ \mathbf{e}_{z} \times \mathbf{k}_{\perp}\right] \bm{\hat{\sigma}} $, which is called Rashba SOI where the parameter $ \lambda_{BR} \propto \mathcal{E} $ is Bychkov-Rashba constant. For a long time this SOI  (see Eq. (\ref{H_nrl})) \textit{a priori} was considered as the only one and as the one appropriate for all physical situations.  

Nevertheless, as it has been shown in Refs. \cite{AoP1,AoP2}, there can exist several possible solutions of the DE which are different from each other and correspond to different spin states of a particle.  More precisely, they correspond to different mutual directions of the spin quantization axis and the direction of the momentum. It has turned out that there is a finite number of various situations, among them also the situation which is different from the Rashba case, the difference between which is controlled by the so called spin invariants. These invariants commute with the Dirac Hamiltonian but do not commute between themselves  \cite{Sokolov}. This fact has allowed to obtain the general solution of the DE and to calculate all relativistic corrections to it in a full agreement with the analytical general solution.   In these papers, however, the SchE which includes such corrections has not been derived. In the present paper we find the explicit form for SOI terms in the SchE which lead to the same energy corrections that follow from the exact solution found in \cite{FNT}. 

The paper is organized as follows. In Section 2 the essential information of the quantum field  theory which allows to write down the Dirac Hamiltonian in the second quantization representation, is given. In Section 3 it is shown that with the accuracy of the second order of the ratio $ \vert V \vert /mc^{2} \ll 1 $ the Hamiltonian of particles and antiparticles, linked via external potential field $ V(\mathbf{r}) $, can be transformed to the Hamiltonian of non-interacting electrons and positrons as it takes place in the case of free particles. In Section 4 we describe the operator invariants controlling spin states of relativistic particles. Section 5 deals with the transition in the electron Hamiltonian to the non-relativistic limit with account of all relativistic corrections among which there are the new ones. The case of the potential in the form of the QW is considered in Section 6 using the results of the previous section. In particular, the Hamiltonian is derived which describes the states of 2D electrons. Their states corresponding to the basic spin invariants and to the generalized invariant, are studied. It is also analyzed when and how in the general approach the spin-orbit band splitting arises in the Rashba or Dresselhaus form.

%\section{Relativistic Hamiltonian} \label{RelHam}
\section{2. Relativistic Hamiltonian}

Let us start with the DE for a particle in the external field: 
$$
i\hbar \frac{\partial \Psi}{\partial t} = \left[ c \left( \hat{\mathbf{p}} - \frac{e}{c} \mathbf{A} \right) \bm{\hat{\alpha}} + V(\mathbf{r}) \hat{I} + m c^{2} \hat{\beta}  \right] \Psi 
$$ 
where $ e $ is elementary charge, $ \mathbf{A} $ is vector-potential of the external electromagnetic field, $ V (\mathbf{r}) $, similar to (\ref{H_nrl}), is the potential, $ \bm{\hat{\alpha}} = \sum_{j} \mathbf{e}_{j} \hat{\alpha}_{j} $ is vector matrix whose components $ \hat{\alpha}_{j} $ ($ j=x,y,z $) together with the unit matrix $ \hat{I} $ and matrix $ \hat{\beta} $ are hermitian Dirac matrices (DM), and, finally, $ \Psi (t;\mathbf{r}) = \left( \psi_{1} \: \psi_{2} \: \psi_{3} \: \psi_{4} \right)^{T} $ is a 4-component function, known also as a bispinor, or 4-spinor (here and below a symbol «$ \wedge $» ('hat') is used over matrices and matrix operators, only).

According to the quantum field theory, the DE is the Euler-Lagrange equation which follows from the variation of the Lagrange functional density $ \mathcal{L} $. It depends on the two 4-component variables, bispinors $ \Psi $, namely, on its components $ \psi_{\mu} $ ($ \mu = 1,2,3,4 $), and $ \bar{\Psi} = \Psi^{\dagger} \hat{\beta} $ is the Dirac conjugated bispinor (see, e.g., Refs. \cite{Bete,RelQuant}). The transition to the Hamilton form is provided by introducing generalized momenta, 
\[
\mathcal{P} = \frac{\partial \mathcal{L}}{\partial \dot{\Psi}} = i \hbar \bar{\Psi}\hat{\beta} = i\hbar \Psi^{\dagger} , \quad \dot{\Psi} = \frac{\partial \Psi}{\partial t}, 
\]
which are canonically conjugate to the components of the bispinor $ \Psi $, and Hamilton functional density, 
\[
\mathcal{H} = i\hbar \Psi^{\dagger} \frac{\partial \Psi}{\partial t} - \mathcal{L} = \Psi^{\dagger}(\mathbf{r},t) \hat{H}_{D} \Psi (\mathbf{r},t) ,
\]
in which the operator
\begin{equation}
\label{H_D} 
\hat{H}_{D} = c \left( \hat{\mathbf{p}} - \frac{e}{c} \mathbf{A} \right) \bm{\hat{\alpha}} + V(\mathbf{r}) \hat{I} + m c^{2} \hat{\beta} 
\end{equation} 
is the Dirac Hamiltonian. Subsequently, the spinor field operator is reduced to the integral  
\[
\mathrm{H} = \int \mathcal{H} d\mathbf{r} = \int \Psi^{\dagger}(\mathbf{r}) \hat{H}_{D} \Psi (\mathbf{r}) d\mathbf{r} 
\]
where the spatial integration is carried out over the whole volume. Here the bispinor $ \Psi $ is considered not as the Schr{\"o}dinger wave function, but as an amplitude of some physical field which is called 'a spinor field' whose components are $ q $-numbers in the meaning that the inequality $ \psi_{\nu}^{\ast} \psi_{\mu} \neq \psi_{\mu}\psi_{\nu}^{\ast} $ takes place. 

Consider, first, particle dynamics in the absence of the magnetic field, $ \mathbf{A} = 0 $. The Dirac Hamiltonian, (\ref{H_D}), can be represented in the form of the sum of two terms $ \hat{H}_{D} = \hat{H}_{D}^{(0)} + \hat{V}_{D} $, one of which is the Hamiltonian of a free particle,  
\begin{equation}
\label{mH_D(0)+mV} 
\hat{H}_{D}^{(0)} = c \hat{\mathbf{p}} \bm{\hat{\alpha}} + m c^{2}\hat{\beta} , 
\end{equation}
and the second one accounts for particle interaction with the external field, $ \hat{V}_{D} = V(\mathbf{r}) \hat{I} $. This transforms the spinor field Hamiltonian to the form: 
\begin{equation}
\label{H_0+V} 
\mathrm{H} = \int \left( \Psi^{\dagger}(\mathbf{r}) \hat{H}_{D}^{(0)} \Psi(\mathbf{r}) + \Psi^{\dagger}(\mathbf{r}) V(\mathbf{r}) \Psi(\mathbf{r}) \right) d\mathbf{r} .
\end{equation}
Any bispinor, $ \Psi(\mathbf{r}) $, can be expanded over the complete ortho-normalized bispinor system, in particular, the one for free particles, i.e., bispinors which satisfy the equation $ \hat{H}_{D}^{(0)} \Psi^{(0)}(\mathbf{r}) = E \Psi^{(0)}(\mathbf{r}) $.

Since a free particle momentum is conserved, it is convenient to undertake the transition to the momentum representation and to use Fourier components of the eigen bispinor $ \Psi^{(0)} $, i.e., to expand bispinor $ \Psi(\mathbf{r}) $ over plane waves in the cube with the side  $ L $ ($ L \rightarrow \infty $):
\begin{equation}
\label{repr_k}  
\Psi (\mathbf{r}) = \frac{1}{L^{3/2}} \sum_{\mathbf{k}} e^{i\mathbf{k} \mathbf{r}} \Psi^{(0)}\left(\mathbf{k} \right)   .
\end{equation}
Here $ \mathbf{k} = \sum_{j} k_{j}\mathbf{e}_{j} $ with $ k_{j} = (2\pi /L )n_{j} $ ($ j = x,y,z $) where the integer numbers $ n_{j} $ take values from $ -\infty $ to $ \infty $.  In such presentation momentum operator is a $ c $-number (wave vector), $ \mathbf{\hat{p}} \Rightarrow \hbar \mathbf{k} $, and the bispinor components $ \Psi^{(0)}\left(\mathbf{k} \right) $ are determined from the equation 
\begin{equation}
\label{H_0-k} 
\hat{H}^{(0)}_{D}(\mathbf{k}) \Psi^{(0)}\left(\mathbf{k} \right) = E \Psi^{(0)}\left(\mathbf{k} \right) , \quad \hat{H}^{(0)}_{D}(\mathbf{k}) =  \hbar c\mathbf{k} \bm{\hat{\alpha}} + m c^{2}\hat{\beta}  .
\end{equation}

In a block form the 4-line matrices are expressed via the 2-line ones \cite{Bete,RelQuant} 
\begin{equation}
\label{MD} 
\bm{\hat{\alpha}} = \left( 
\begin{array}{cc}
0 & \bm{\hat{\sigma}} \\ \bm{\hat{\sigma}} & 0
\end{array} \right) \, , \quad  \hat{\beta} = \left( \begin{array}{cc}
\hat{I}_{2} & 0 \\ 0 & -\hat{I}_{2}
\end{array}  \right) \,,
\end{equation}
where $ \bm{\hat{\sigma}} $ is the vector operator, whose components are given by the Pauli matrices, and $ \hat{I}_{2} $ is a unit matrix of the second order. It is convenient also to write the bispinor in the block-form, too: $ \Psi^{(0)} \left( \mathbf{k} \right) = \left( \psi_{u}  \left( \mathbf{k} \right) \,\psi_{d} \left( \mathbf{k} \right) \right)^{T} $, where $ \psi_{u} = \left( \psi_{1} \,\psi_{2} \right)^{T} $ and $ \psi_{d} = \left( \psi_{3} \,\psi_{4} \right)^{T} $ are the upper and lower spinors of the bispinor, respectively. Within this scheme Eq. (\ref{H_0-k}) takes a simple form, 
\begin{equation}
\label{H_D_m}
\left( \begin{array}{cc}
m c^{2} \hat{I}_{2} & \hbar c \mathbf{k}\bm{\hat{\sigma}} \\ \hbar c \mathbf{k}\bm{\hat{\sigma}} & -m c^{2} \hat{I}_{2}
\end{array} \right) \left( \begin{array}{c}
\psi_{u} \\
\psi_{d}
\end{array} \right) = E \left( \begin{array}{c}
\psi_{u} \\
\psi_{d}
\end{array} \right) .
\end{equation} 
Its solution can be found, in particular, using the well-known Foldy–Wouthuysen (FW) unitary  transform (see, e.g., Refs. \cite{Davydov,Messia}). The following four ortho-normalized eigen bispinors are the solutions of Eq. (\ref{H_D_m}) (or Eq. (\ref{H_0-k})): 
\begin{equation}
\label{solutn} 
\begin{array}{cc}
\Psi^{(0)}_{p,\sigma}\left( \mathbf{k}\right) = A_{\mathbf{k}} \left( \begin{array}{c}
\chi_{p,\sigma} \\ \frac{\hbar c\mathbf{k}\cdot \bm{\hat{\sigma}}}{\varepsilon \left(\mathbf{k} \right) + mc^{2}} \chi_{p,\sigma} 
\end{array} \right) , &  E = E_{p} \left(\mathbf{k} \right) \equiv \varepsilon \left(\mathbf{k} \right) , \\
\Psi^{(0)}_{a,\sigma}\left( \mathbf{k}\right) = A_{\mathbf{k}} \left( \begin{array}{c}
-\frac{\hbar c\mathbf{k}\cdot \bm{\hat{\sigma}}}{\varepsilon \left(\mathbf{k} \right) + mc^{2}} \chi_{a,\sigma} \\ \chi_{a,\sigma}
\end{array}  \right) , & E = E_{a} \left(\mathbf{k} \right) \equiv - \varepsilon \left(\mathbf{k} \right) , 
\end{array}
\quad A_{\mathbf{k}} = \sqrt{\frac{\varepsilon \left(\mathbf{k} \right) + mc^{2}}{2 \varepsilon \left(\mathbf{k} \right)}} . 
\end{equation}
where $ A_{\mathbf{k}} $ is the normalization coefficient, and  
\begin{equation}
\label{eps(k)} 
\varepsilon \left(\mathbf{k} \right) = \sqrt{m^{2}c^{4} + c^{2}\hbar^{2} \mathbf{k}^{2}} .
\end{equation} 
The spinors $ \chi_{\nu ,\sigma} $ ($ \nu = p,a $) are not fully determined because the bispinors (\ref{solutn}) satisfy Eq. (\ref{H_D_m}) at arbitrary spinors.  

The bispinor $ \Psi^{(0)}_{p,\sigma} $ in Eq. (\ref{solutn}) corresponds to the positive eigen value $  E_{p}  $, and bispinor $ \Psi^{(0)}_{a,\sigma} $ -- to the negative one $ E_{a} < 0 $, which are degenerate. The number $ \sigma $ in Eq. (\ref{solutn}) takes two values, which are assigned to the two arbitrary chosen spinors of the degenerate state. Therefore, the four eigen bispinors (\ref{solutn}) $ \Psi^{(0)}_{\nu,\sigma} $, where each index, $\nu $  and $ \sigma $, takes two values, form a complete ortho-normalized system. The condition of their ortho-normalization $ \left( \Psi^{(0)}_{\nu,\sigma}\right)^{\dagger} \Psi^{(0)}_{\nu',\sigma'} = \delta_{\nu ,\nu'} \delta_{\sigma ,\sigma'} $ directly leads to the ortho-normalization of the corresponding pair of the spinors  $ \chi_{\nu,\sigma}^{\dagger} \chi_{\nu,\sigma'} = \delta_{\sigma ,\sigma'} $. Index $ \sigma $ has the meaning of the spin number and can be assigned the values $ \sigma = \pm 1 $ or $ \sigma = \uparrow ,\downarrow $. The relation between the two ortho-normalized spinors with different values of $ \sigma $ is given by the Kramers relation: $ \chi_{\nu ,-\sigma} = \hat{K}_{K} \chi_{\nu ,\sigma} $, where $ \hat{K}_{K} = - i \hat{\sigma}_{y} K $ is the Kramers operator, which includes $ K $-operation of the complex conjugation. 

This consideration shows the principal possibility of expanding any bispinor $ \Psi \left( \mathbf{r} \right) $ over the bispinors (\ref{solutn}): 
\begin{equation}
\label{finPhi} 
\Psi \left( \mathbf{r} \right) = \frac{1}{L^{3/2}} \sum_{\mathbf{k} ,\sigma} e^{i\mathbf{k} \mathbf{r}} \left( a_{\mathbf{k},\sigma} \Psi^{(0)}_{p,\sigma}\left( \mathbf{k}\right) + b^{\dagger}_{-\mathbf{k},\sigma} \Psi^{(0)}_{a,\sigma}\left( \mathbf{k}\right) \right)   .
\end{equation}
Here we have used the notations accepted in quantum field theory for the creation and annihilation operators, $ a^{\dagger}_{\mathbf{k},\sigma}/ b^{\dagger}_{\mathbf{k},\sigma}$,  $ a_{\mathbf{k},\sigma}/ b_{\mathbf{k},\sigma}$ of a particle/antiparticle with the wave vector $ \mathbf{k} $ and spin number $ \sigma $, respectively. The physical requirement of positive eigen energy values of the Hamiltonian (\ref{H_0+V}) determines Fermi commutation rules (see, e.g., Refs. \cite{RelQuant}):
\[
a_{\mathbf{k},\sigma} a^{\dagger}_{\mathbf{k},\sigma} + a^{\dagger}_{\mathbf{k},\sigma} a_{\mathbf{k},\sigma} = 1 , \quad b_{\mathbf{k},\sigma} b^{\dagger}_{\mathbf{k},\sigma} + b^{\dagger}_{\mathbf{k},\sigma} b_{\mathbf{k},\sigma} = 1  ,
\]
with all other pairs of these operators mutually anti-commuting. 

Substituting expression (\ref{finPhi}) in Eq. (\ref{H_0+V}) and taking into account that 
\[
\frac{1}{L^{3}} \int_{-L/2}^{L/2} \exp \left[ i(\mathbf{k}_{1} - \mathbf{k}_{2})\mathbf{r}\right] d\mathbf{r} = \delta_{\mathbf{k}_{1},\mathbf{k}_{2}} \equiv \delta_{k_{1x},k_{2x}} \delta_{k_{1y},k_{2y}} \delta_{k_{1z},k_{2z}}, 
\]
where $ \delta_{\mathbf{k}_{1},\mathbf{k}_{2}} $ is Kronecker symbol, one derives the Hamiltonian in the occupation number representation, 
\begin{equation}
\label{H_rel1} 
\begin{array}{c}
\mathrm{H} = \sum_{\mathbf{k},\sigma} \left[ \varepsilon(\mathbf{k}) a_{\mathbf{k},\sigma}^{\dagger} a_{\mathbf{k},\sigma} + \frac{1}{L^{3}} \sum_{\mathbf{k}',\sigma'} \mathcal{V}_{\sigma ,\sigma'}^{(p)} \left(\mathbf{k}, \mathbf{k}' \right) a_{\mathbf{k},\sigma}^{\dagger} a_{\mathbf{k}',\sigma'} - \right. \\
- \varepsilon(\mathbf{k}) b_{-\mathbf{k},\sigma} b^{\dagger}_{-\mathbf{k},\sigma} + \frac{1}{L^{3}} \sum_{\mathbf{k}',\sigma'} \mathcal{V}_{\sigma ,\sigma'}^{(a)}\left(\mathbf{k}, \mathbf{k}' \right) b_{-\mathbf{k},\sigma} b^{\dagger}_{-\mathbf{k}',\sigma'} + \\
\left. + \frac{1}{L^{3}} \sum_{\mathbf{k}',\sigma'}  \left( \mathcal{V}_{\sigma ,\sigma'}^{(p-a)}\left(\mathbf{k}, \mathbf{k}' \right) a^{\dagger}_{\mathbf{k},\sigma} b^{\dagger}_{-\mathbf{k}',\sigma'} + \mathcal{V}_{\sigma ,\sigma'}^{(a-p)} \left(\mathbf{k}, \mathbf{k}' \right) b_{-\mathbf{k},\sigma} a_{\mathbf{k}',\sigma'} \right) \right]  ,
\end{array}
\end{equation} 
where 
\begin{equation}
\label{calV_p,a} 
\begin{array}{c}
\mathcal{V}_{\sigma ,\sigma'}^{(\nu)}\left(\mathbf{k}, \mathbf{k}' \right) = V \left(\mathbf{k} - \mathbf{k}' \right) \left( \Psi^{(0)}_{\nu,\sigma}\left( \mathbf{k}\right) \right)^{\dagger} \Psi^{(0)}_{\nu,\sigma'}\left( \mathbf{k}'\right) , \quad \nu = p,a ; \\
\mathcal{V}_{\sigma ,\sigma'}^{(p-a)}\left(\mathbf{k}, \mathbf{k}' \right) = V \left(\mathbf{k} - \mathbf{k}' \right) \left( \Psi^{(0)}_{p,\sigma}\left( \mathbf{k}\right) \right)^{\dagger} \Psi^{(0)}_{a,\sigma'}\left( \mathbf{k}'\right) ,\\
\mathcal{V}_{\sigma ,\sigma'}^{(a-p)}\left(\mathbf{k}, \mathbf{k}' \right) = \left(  \mathcal{V}_{\sigma' ,\sigma}^{(p-a)}\left(\mathbf{k}', \mathbf{k} \right)\right)^{\ast} , \quad  V\left(\mathbf{k} \right) = \int e^{-i\mathbf{k} \mathbf{r}} V(\mathbf{r}) d\mathbf{r} .
\end{array}
\end{equation}

From the above it follows that the operator (\ref{H_rel1}) is the sum of the three terms, $ \mathrm{H} = \mathrm{H}_{p} + \mathrm{H}_{a} + \mathrm{V}_{p-a} $, which are the Hamiltonians of particles, $ \mathrm{H}_{p} $, antiparticles, $ \mathrm{H}_{a} $, and operator $ \mathrm{V}_{p-a} $ which describes their mutual transformation under the scattering in the external potential. After transforming the product of creation and annihilation operators to the standard form, the Hamiltonian (\ref{H_rel1}) becomes positively determined, except, according to quantum field theory postulates \cite{RelQuant}, the infinite additive constant, i.e., energy of the state in the absence of any particles, vacuum state, from which energies of all elementary excitations of the spinor field, particles and antiparticles, are calculated.

The operator $ \mathrm{V}_{p-a} $ describes mixing of particle and antiparticle states in the external field, and necessarily has to be taken into account. The transformation which separates particle and antiparticle states in the Hamiltonian (\ref{H_rel1}) exactly, is not known. Nevertheless, in the case of interaction (\ref{calV_p,a}) the perturbation theory can be used with any required accuracy. 

%\section{Approximate renormalization \label{ApprRenorm}}
\section{3. Approximate renormalization}

An important problems of non-relativistic physics is approximate separation of particle and antiparticle states. This is possible for potentials that satisfy the inequality $ \vert V \left(\mathbf{r} \right) \vert \ll mc^{2} $. In such a case the operator $ \mathrm{V}_{p-a} $ can be considered as a perturbation and one can use the canonical transformation of the Hamiltonian $ \mathrm{H} = \mathrm{H}_{0} + \lambda \mathrm{V} $ with parameter $ \lambda $ characterizing the smallness of the perturbation, to the new representation $ \tilde{\mathrm{H}} = \exp (-\lambda S) \mathrm{H} \exp (\lambda S) $ in which the value of the non-diagonal part of the Hamiltonian exceeds the given accuracy. For small perturbations,  $ \lambda \ll 1 $, the exponent can be expanded into the series with respect to the degrees of $ \lambda $, and the Hamiltonian after the Schrieffer-Wolff transformation \cite{Schr-W} takes the form 
\[
\tilde{\mathrm{H}} = e^{-\lambda S} \mathrm{H} e^{\lambda S} = \mathrm{H} + \lambda \left[ \mathrm{H},S \right] + \frac{\lambda^{2}}{2} \left[ \left[ \mathrm{H},S \right] S \right] + \ldots \, .
\]

To diagonalize this Hamiltonian up to the given accuracy, e.g., up to $ \lambda^{n} $, it is convenient to search the operator $ S $ ($= -S^{\dagger} $) in the form $ \lambda S = \sum_{j =1}^{n} \lambda^{j} S_{j} $. In particular, for diagonalization up to the second order $ \sim \lambda^{2} $, it is enough to preserve  only the first term in the operator $ \lambda S $ and to choose it from the condition 
\[
\mathrm{V} + \left[ \mathrm{H}_{0},S_{1} \right] = 0 .
\]
Taking into account the explicit form of the operator $ \mathrm{V} = \mathrm{V}_{p-a} $ in the Hamiltonian (\ref{H_rel1}), we get 
\begin{equation}
\label{S} 
\begin{array}{cc}
S_{1} = & L^{-3} \sum_{\mathbf{k},\mathbf{k}',\sigma ,\sigma'} \frac{ V \left(\mathbf{k} - \mathbf{k}' \right)}{\varepsilon(\mathbf{k}) + \varepsilon(\mathbf{k}')} \left[ \left( \Psi^{(0)}_{a,\sigma}\left( \mathbf{k}\right) \right)^{\dagger} \Psi^{(0)}_{p,\sigma'}\left( \mathbf{k}'\right) b_{-\mathbf{k},\sigma} a_{\mathbf{k}',\sigma'} - \right. \\
 & \left. - \left( \Psi^{(0)}_{p,\sigma}\left( \mathbf{k}\right) \right)^{\dagger} \Psi^{(0)}_{a,\sigma'}\left( \mathbf{k}'\right) a^{\dagger}_{\mathbf{k},\sigma} b^{\dagger}_{-\mathbf{k}',\sigma'} \right]  .
\end{array}
\end{equation}
From the last expression it follows that the renormalization with the second order accuracy is correct if the inequality    
\begin{equation}
\label{condV} 
\frac{ \mid V \mid}{2 m c^{2}} \ll 1 , \quad \mid V \mid = V_{max} - V_{min}.
\end{equation} 
is fulfilled. It, as a rule, is valid for any non-relativistic potential.  
 
Therefore, up to terms $ \sim\lambda^{2} $ the operator (\ref{H_rel1}) reduces to the form
\[
\tilde{\mathrm{H}} = \mathrm{H}_{0} + \frac{\lambda^{2}}{2} \left[ \mathrm{H}_{1},S_{1} \right] + \ldots = \tilde{\mathrm{H}}_{p} + \tilde{\mathrm{H}}_{a} + \mathit{O} (\lambda^{4}) ,
\]
in which the states of particles and antiparticles turn out to be independent and can be considered using the Hamiltonians 
\begin{equation}
\label{tildeH_p} 
\tilde{\mathrm{H}}_{p} = \sum_{\mathbf{k},\sigma} \left[ \varepsilon(\mathbf{k}) a_{\mathbf{k},\sigma}^{\dagger} a_{\mathbf{k},\sigma} + \frac{1}{L^{3}} \sum_{\mathbf{k}',\sigma'} \left( \mathcal{V}_{\sigma ,\sigma'}^{(p)} \left(\mathbf{k}, \mathbf{k}' \right) + W^{(p)}_{\sigma ,\sigma'} \left( \mathbf{k} ,\mathbf{k}'\right)\right) a_{\mathbf{k},\sigma}^{\dagger} a_{\mathbf{k}',\sigma'} \right]   
\end{equation}
\begin{equation}
\label{tildeH_a} 
\tilde{\mathrm{H}}_{a} = \sum_{\mathbf{k},\sigma} \left[ \varepsilon(\mathbf{k})  b^{\dagger}_{\mathbf{k},\sigma} b_{\mathbf{k},\sigma} - \frac{1}{L^{3}} \sum_{\mathbf{k}',\sigma'} \left( \mathcal{V}_{\sigma ,\sigma'}^{(a)} \left(\mathbf{k}, \mathbf{k}' \right) - W^{(a)}_{\sigma' ,\sigma} \left( -\mathbf{k}' ,-\mathbf{k}\right) \right)  b^{\dagger}_{\mathbf{k},\sigma} b_{\mathbf{k}',\sigma'} \right]    
\end{equation}
for particles and antiparticles, respectively. In the above equations the amplitudes $ \mathcal{V}_{\sigma ,\sigma'}^{(\nu)} \left(\mathbf{k}, \mathbf{k}' \right) $ are given by the expressions (\ref{calV_p,a}) and notations 
\begin{equation}
\label{W_p,a} 
\begin{array}{cc}
W^{(p)}_{\sigma ,\sigma'} \left( \mathbf{k} ,\mathbf{k}'\right) = \frac{1}{2L^{3}} \sum_{\mathbf{k}_{1},\sigma_{1} } V \left(\mathbf{k} - \mathbf{k}_{1} \right) V \left(\mathbf{k}_{1} - \mathbf{k}' \right) \frac{ \varepsilon(\mathbf{k}) + 2\varepsilon(\mathbf{k}_{1}) + \varepsilon(\mathbf{k}') }{\left[ \varepsilon(\mathbf{k}) + \varepsilon(\mathbf{k}_{1}) \right] \left[ \varepsilon(\mathbf{k}_{1}) + \varepsilon(\mathbf{k}') \right] } \times \\
 \left( \Psi^{(0)}_{p,\sigma}\left( \mathbf{k}\right) \right)^{\dagger} \Psi^{(0)}_{a,\sigma_{1}}\left( \mathbf{k}_{1}\right) \left( \Psi^{(0)}_{a,\sigma_{1}} \left( \mathbf{k}_{1}\right) \right)^{\dagger} \Psi^{(0)}_{p,\sigma'}\left( \mathbf{k}'\right) , \\
W^{(a)}_{\sigma ,\sigma'} \left( \mathbf{k} ,\mathbf{k}'\right) = W^{(p \rightarrow a)}_{\sigma ,\sigma'} \left( \mathbf{k} ,\mathbf{k}'\right) 
\end{array}
\end{equation}
are used. In the Hamiltonian $ \tilde{\mathrm{H}}_{a} $, the product of creation and annihilation operators is written in the normal form with the change $ \mathbf{k} \rightarrow - \mathbf{k} $ in the sum, and the energy of the vacuum is deduced. 

The creation and annihilation operators of particles/antiparticles in the expansion (\ref{finPhi}) have the indeces $ \mathbf{k} $ and $ \sigma $, which have the meaning of the quantum numbers with $ \sigma $ corresponding to the spinor $ \chi_{p/a,\sigma} $ in bispinors (\ref{solutn}), while the spinors themselves are not determined uniquely. Their arbitrariness in Eq. (\ref{solutn}) indicates that in the general case the Hamiltonian (\ref{H_rel1}) is 'invariant' with respect to the choice of spinors. In particular, without loss of generality, one can choose the following pair of the orthogonal spinors $ \chi_{\uparrow} = \left( 1 \, 0 \right)^{T} $, $ \chi_{\downarrow} = \left( 0 \, 1 \right)^{T} $, i.e., to use the simplest spin functions related to the initial coordinate system, which are eigen functions of the operator $ \hat{\sigma}_{z} $. 

For free particles such choice is not important, since the energy is independent of the spin variable, and any spinor can be expressed via the two other spinors. Therefore, one can use solutions (\ref{solutn}) with spinors $ \chi_{p/a,\sigma} $ which can be chosen according to convenience.  

Below we shall show that spin polarization in concrete spin states is tightly related to the wave vector, characterizing spatial motion. This relation is described by some new effective interaction called SOI. According to expressions (\ref{calV_p,a}) and (\ref{W_p,a}), this interaction is directly determined not only by the symmetry of the field in which particle propagates, but also by its spin when the particle is scattered by this field (see the Hamiltonian (\ref{H_rel1}) and expressions (\ref{tildeH_p})-(\ref{tildeH_a})).

%\section{ Spin invariants \label{SpinInv}}
\section{4. Spin invariants}

As it has been reminded above, particle spin states, i.e., concrete form of the spinors in Eq. (\ref{solutn}), can be found using spin invariants. For example, the existence of these invariants has allowed to find new spin states of quasi-2D electrons \cite{AoP1,AoP2}. When a particle is free, its dynamics can be characterized by several invariants \cite{Sokolov}: vector of \textit{magnetic spin polarization},
\begin{equation}
\label{mu}
\bm{\hat{\mu}} = \bm{\hat{\Sigma}} + \frac{1}{mc} \bm{\hat{\Gamma}} \times \mathbf{\hat{p}} = \bm{\hat{\Sigma}} - \frac{1}{mc} \hbar \mathbf{k} \times  \bm{\hat{\Gamma}} \equiv \bm{\hat{\mathcal{I}}}^{(\bm{\mu})}\left( \mathbf{k} \right), 
\end{equation} 
vector of \textit{spin polarization},  
\begin{equation}
\label{genS}
\bm{\hat{\mathcal{S}}} = \bm{\hat{\Omega}} + \hat{\rho}_{1} \frac{\mathbf{\hat{p}}}{mc}  = \bm{\hat{\Omega}} + \frac{\hbar \mathbf{k}}{mc} \hat{\rho}_{1} \equiv \bm{\hat{\mathcal{I}}}^{(\bm{\mathcal{S}})}\left( \mathbf{k} \right),
\end{equation}
\textit{helicity} $ \hat{h} = \mathbf{\hat{p}} \bm{\hat{\Sigma}} $, and vector of \textit{electric spin polarization } $ \bm{\hat{\epsilon}} = - \mathbf{\hat{p}} \times \bm{\hat{\Omega}} $. It is easy to see that the latter two invariants can be represented in the form $ \hat{h} = \mathbf{\hat{p}} \bm{\hat{\mu}} $ and $ \bm{\hat{\epsilon}} = - \mathbf{\hat{p}} \times \bm{\hat{\mathcal{S}}} $, respectively. In formulas (\ref{mu}) and (\ref{genS}) we have taken into account that  $ \mathbf{\hat{p}} \Rightarrow \mathbf{p} = \hbar \mathbf{k} $. 

The three vector DMs, according to (\ref{MD}), have the block form  
\begin{equation}
\label{mS,G,O,r} 
\bm{\hat{\Sigma}} = \left( 
\begin{array}{cc}
\bm{\hat{\sigma}} & 0 \\ 0 & \bm{\hat{\sigma}}
\end{array} \right)  , \; \bm{\hat{\Gamma}} = \left( 
\begin{array}{cc}
0 & -i\bm{\hat{\sigma}} \\ i\bm{\hat{\sigma}} & 0
\end{array} \right) , \; \bm{\hat{\Omega}} = \left( 
\begin{array}{cc}
\bm{\hat{\sigma}} & 0 \\ 0 & - \bm{\hat{\sigma}}
\end{array} \right) , \; \hat{\rho }_{1} = \left( \begin{array}{cc}
0 & \hat{I}_{2} \\ \hat{I}_{2} & 0
\end{array}  \right) .
\end{equation}
The matrix $ \bm{\hat{\Sigma}} $ determines the spin operator $ \mathbf{\hat{s}} = (\hbar /2) \bm{\hat{\Sigma}} $ \cite{Bete,RelQuant,Messia}, and its projections, $ \hat{\Sigma}_{j} = - i \hat{\alpha}_{k}\hat{\alpha}_{l} $, have different spatial indeces, which form a cyclic permutation. The other projections are given by the expressions $ \hat{\Gamma}_{j} = -i \hat{\beta} \hat{\alpha}_{j} $, $ \hat{\Omega}_{j} = \hat{\beta} \hat{\Sigma}_{j} $, and, finally, the DM $ \hat{\rho }_{1} = -i \hat{\alpha}_{j} \hat{\alpha}_{k} \hat{\alpha}_{l} $ has the same three cyclic indeces.  

Since the invariants commute with the Hamiltonian, $ \hat{H}^{(0)}_{D} $, their eigen bispinors are compatible with the eigen bispinors of Eq. (\ref{H_0-k}), or matrix equation (\ref{H_D_m}). This means that the bispinors (\ref{solutn}) satisfy the equalities 
\[
\hat{\mathcal{I}}^{(j)}\left( \mathbf{k} \right) \Psi^{(0)}_{\nu,\sigma}\left( \mathbf{k}\right) = s_{j,\nu ,\sigma }\left( \mathbf{k} \right) \Psi^{(0)}_{\nu,\sigma}\left( \mathbf{k}\right) ,
\]
where $ \hat{\mathcal{I}}^{(j)}(\mathbf{k}) $ is one of the invariants (including Cartesian projections in the case of vector matrices), and $ s_{j,\nu ,\sigma}\left( \mathbf{k} \right) $ is the corresponding eigen value. Namely this equality defines the concrete form of the spinors $ \chi_{\nu,\sigma} $ in (\ref{solutn}). As the operators (\ref{mu})-(\ref{genS}) do not commute, each of them corresponds to its own pair of eigen spinors. 

Worth mentioning, any arbitrary linear combination of these invariants also commutes with the Hamiltonian $ \hat{H}^{(0)}_{D} $ and can be considered as some generalized invariant 
\begin{equation}
\label{Inv_g2} 
\hat{\mathcal{I}}_{gen} = \mathbf{r}_{\bm{\mu}}\left( \mathbf{k} \right) \bm{\hat{\mu}} + \mathbf{r}_{\bm{\mathcal{S}}} \left( \mathbf{k} \right) \bm{\hat{\mathcal{S}}} . 
\end{equation}
Then, choosing $ \mathbf{r}_{\bm{\mu}}\left( \mathbf{k} \right) = \mathbf{r}_{\bm{\mu}} $, $ \mathbf{r}_{\bm{\mathcal{S}}} = 0 $ or $ \mathbf{r}_{\bm{\mu}}\left( \mathbf{k} \right) = 0 $, $ \mathbf{r}_{\bm{\mathcal{S}}}\left( \mathbf{k} \right) = \mathbf{r}_{\bm{\mathcal{S}}} $, one gets both expressions (\ref{mu}) and (\ref{genS}), respectively. A possible dependence of the coefficients $ \mathbf{r}_{\bm{\mu}}\left( \mathbf{k} \right) $ and $ \mathbf{r}_{\bm{\mathcal{S}}}\left( \mathbf{k} \right)$ on the momentum $ \mathbf{p} = \hbar \mathbf{k} $ can be used to find the two other invariants. Indeed, if $ \mathbf{r}_{\bm{\mu}} \left( \mathbf{k} \right) \sim \mathbf{k} $, and $ \mathbf{r}_{\bm{\mathcal{S}}} = 0 $, one can get from (\ref{Inv_g2}) $ \hat{\mathcal{I}}_{gen} = \hbar \mathbf{k} \bm{\hat{\mu}} = \hbar \mathbf{k} \bm{\hat{\Sigma}} \equiv \hat{h} $; if $ \mathbf{r}_{\bm{\mu}}\left( \mathbf{k} \right) = 0 $, and $ \mathbf{r}_{\bm{\mathcal{S}}} \left( \mathbf{k} \right) \sim \mathbf{k} \times \mathbf{e}_{j} $, $ \hat{\mathcal{I}}_{gen} \sim \hat{\epsilon}_{j} $, which reconstructs $ j $-projection of the operator $ \bm{\hat{\epsilon}} $. Therefore, choosing various coefficients in the operator (\ref{Inv_g2}), we can represent any invariants and their linear combinations. 

Using the expansion (\ref{finPhi}) and the explicit forms of the bispinors (\ref{solutn}) and matrices (\ref{mS,G,O,r}), we come to the following expression for the invariant  (\ref{Inv_g2}): 
\begin{equation}
\label{op-Inv} 
\mathfrak{I}_{gen} = \int \Psi^{\dagger} \hat{\mathcal{I}}_{gen} \Psi d\mathbf{r} = \sum_{\mathbf{k},\sigma , \sigma'} \left( \chi_{p,\sigma}^{\dagger} \mathbf{r}_{p} \bm{\hat{\sigma}} \chi_{p,\sigma'} a_{\mathbf{k},\sigma}^{\dagger} a_{\mathbf{k},\sigma'} + \chi_{a,\sigma}^{\dagger} \mathbf{r}_{a} \bm{\hat{\sigma}} \chi_{a,\sigma'} b_{-\mathbf{k},\sigma} b_{-\mathbf{k},\sigma'}^{\dagger} \right) ,
\end{equation}
where 
\begin{equation}
\label{r_+,-} 
\mathbf{r}_{p/a} = \left( \frac{\varepsilon(\mathbf{k})}{mc^{2}} \mathbf{r}_{\bm{\mu}}(\mathbf{k}) - \frac{\hbar^{2}c^{2}\left( \mathbf{r}_{\bm{\mu}}\mathbf{k} \right)}{mc^{2} \left[ \varepsilon (\mathbf{k}) + mc^{2}\right]}  \mathbf{k} \right) \pm \left( \mathbf{r}_{\bm{\mathcal{S}}}(\mathbf{k}) + \frac{\hbar^{2}c^{2} \left( \mathbf{r}_{\bm{\mathcal{S}}} \mathbf{k} \right)}{mc^{2} \left[ \varepsilon (\mathbf{k}) + mc^{2}\right]} \mathbf{k} \right) . 
\end{equation}
In view of the fact that the operator  (\ref{op-Inv}) splits into the sum   $ \mathfrak{I}_{gen} = \mathfrak{I}_{gen}^{(p)} + \mathfrak{I}_{gen}^{(a)} $,  the invariants for particles $ \mathfrak{I}_{gen}^{(p)} $ and antiparticles  $  \mathfrak{I}_{gen}^{(a)} $ can be diagonalized with respect to the spin variables independently. This can be performed using the bispinors  $ \chi_{\nu,\sigma} $, which are eigen bispinors of the matrices $ \mathbf{r}_{\nu} \bm{\hat{\sigma}} $, and, hence, satisfy the equations
\begin{equation}
\label{egnspinor_p,a} 
\mathbf{r}_{\nu}  \bm{\hat{\sigma}} \chi_{\nu ,\mathbf{k},\sigma} = \sigma \mid \mathbf{r}_{\nu} \left( \mathbf{k} \right) \mid \chi_{\nu ,\mathbf{k},\sigma} , \quad \nu = (p,a) .
\end{equation}

The direction of the vector $ \mathbf{r}_{p/a} $ determines the quantization axis of the  spin $ \mathbf{\hat{s}}_{p/a} $ of particles/antiparticles and at $ \mathbf{r}_{p} \neq  \mathbf{r}_{a} $ the corresponding axes do not coincide. Moreover, the matrices $ \mathbf{r}_{\nu}  \bm{\hat{\sigma}} $ turn out to be spin invariants of free motion which in the coordinate representation under the change $ \hbar \mathbf{k} \Rightarrow \mathbf{\hat{p}} $ are independent invariants for particles $ \mathfrak{I}_{gen}^{(p)} = \mathbf{r}_{p} \bm{\hat{\sigma}} $ and antiparticles $ \mathfrak{I}_{gen}^{(a)} = \mathbf{r}_{a}  \bm{\hat{\sigma}} $, where $ \mathbf{r}_{\nu}\equiv \mathbf{r}_{\nu} \left( \mathbf{\hat{p}} \right) $. 

In the Cartesian coordinate system arbitrary vectors $ \mathbf{r}_{\bm{\mu}}\left( \mathbf{k} \right) $ and $ \mathbf{r}_{\bm{\mathcal{S}}} \left( \mathbf{k} \right) $ in Eq. (\ref{r_+,-}) can be represented as described  $ \mathbf{r}_{\nu}\left( \mathbf{k} \right) = x_{\nu}\left( \mathbf{k} \right) \mathbf{e}_{x} + y_{\nu}\left( \mathbf{k} \right) \mathbf{e}_{y} + z_{\nu}\left( \mathbf{k} \right) \mathbf{e}_{z} $ ($ \nu = \bm{\bm{\mu}, \mathcal{S}} $), where  $ \mathbf{e}_{j} $ ($ j=x,y,z $) are the basis vectors. This means that the spin polarization of particles is determined by six independent parameters, that are coordinates of these two vectors.  

On the other hand, to describe polarization of free particles which are characterized by the wave vector $ \mathbf{k} $, it is sometimes convenient to make a transition in the momentum space to a local reper which is given by the three mutually orthogonal unit vectors $ \mathbf{e}_{1} $, $ \mathbf{e}_{2} $ and $ \mathbf{e}_{3} $ that are related to a unit vector $ \mathbf{e}_{1} \times \mathbf{e}_{2} = \mathbf{e}_{3} = \mathbf{k} /\vert \mathbf{k} \vert $. It is easy to see that the vectors $ \mathbf{e}_{1} $ and $ \mathbf{e}_{2} $, remaining orthogonal are fixed with the accuracy of rotations around the propagation axis $ \mathbf{e}_{3} \shortparallel \mathbf{k} $. Then each of the vectors (\ref{r_+,-}) has also another expansion, $ \mathbf{r}_{\nu} \left( \mathbf{k} \right) = \xi_{\nu}\left( \mathbf{k} \right) \mathbf{e}_{1} + \eta_{\nu}\left( \mathbf{k} \right) \mathbf{e}_{2} + \zeta_{\nu}\left( \mathbf{k} \right) \mathbf{e}_{3} $. Respectively, vector $ \mathbf{r}_{p} $ in  Eq. (\ref{r_+,-})  which determines  electron spin quantization axis in this local basis, takes the form
\[
\mathbf{r}_{p} = \mathbf{r}_{\bm{\mu}} \left( \mathbf{k} \right) + \mathbf{r}_{\bm{\mathcal{S}}} \left( \mathbf{k} \right) + \frac{\varepsilon (\mathbf{k}) - mc^{2}}{mc^{2}} \left[  \xi_{\bm{\mu}} \left( \mathbf{k} \right) \mathbf{e}_{1} + \eta_{\bm{\mu}} \mathbf{e}_{2} \left( \mathbf{k} \right) + \zeta_{\bm{\mathcal{S}}} \left( \mathbf{k} \right) \mathbf{e}_{3} \right]  .
\]  

Note that in the states corresponding to the invariant $ \mathfrak{I}_{gen}^{(p)} $ with  independent on $ \mathbf{k} $ components  $ \xi_{\nu}\left( \mathbf{k} \right) = \xi_{\nu} $, $ \eta_{\nu} \left( \mathbf{k} \right)= \eta_{\nu} $, $ \zeta_{\nu} \left( \mathbf{k} \right) = \zeta_{\nu} $ ($ \nu = \bm{\bm{\mu}, \mathcal{S}} $) in a local system, among which there are vanishing components,  $ \xi_{\bm{\mu}} = \eta_{\bm{\mu}} = \zeta_{\bm{\mathcal{S}}} = 0 $, spin quantization axis is determined by the vector  $ \mathbf{r}_{p} = \xi_{\bm{\mathcal{S}}} \mathbf{e}_{1} + \eta_{\bm{\mathcal{S}}} \mathbf{e}_{2} + \zeta_{\bm{\mu}} \mathbf{e}_{3} $. This means that in such a basis particle spin orientation is the same for all momenta. In a particular case, when $ \xi_{\bm{\mathcal{S}}} = \eta_{\bm{\mathcal{S}}} = 0 $, $ \zeta_{\bm{\mu}} = 1 $, this corresponds to a spiral state in which particle spin and momentum are parallel. In the case $ \zeta_{\bm{\mu}} = 0 $ and $ \eta_{\bm{\mathcal{S}}} = 1 $ a particle's spin and momentum are orthogonal. Such spin state corresponds to the projection of the operator $ \bm{\hat{\epsilon}} $.  

In the external field the spin operator $ \mathbf{r}_{\nu}\left( \mathbf{k} \right) \bm{\hat{\sigma}} $ can commute with the Hamiltonian (\ref{tildeH_p}) or (\ref{tildeH_a}) only at certain values of the coefficients $ \mathbf{r}_{\bm{\mu}}\left( \mathbf{k} \right) $ and $ \mathbf{r}_{\bm{\mathcal{S}}}\left( \mathbf{k} \right) $ which give the direction of the quantization axis with account of the concrete symmetry of the field. 

%\section{General form of the Hamiltonian with the relativistic corrections \label{GenForm}}

\section{5. General form of the Hamiltonian with the relativistic corrections}

In the previous Section we have used the ratio (\ref{condV}) as a small parameter. In the non-relativistic case the inequality $ \hbar k/mc \ll 1 $ is also valid. Respectively, the Hamiltonians (\ref{tildeH_p}) and (\ref{tildeH_a}) can be expanded using this ratio as a small parameter as well. In this case the kinetic energy can be written as  
\[
\varepsilon(\mathbf{k}) = mc^{2} + \frac{\hbar^{2} k^{2}}{2m} \left( 1 - \frac{\hbar^{2} k^{2}}{4m^{2}c^{2}} \right) .
\]
In view of Eq. (\ref{solutn}), the convolution of the bispinors in Eqs. (\ref{calV_p,a}) and (\ref{W_p,a}) has to be expanded also: 
\begin{equation}
\label{prbisp_nr} 
\begin{array}{c}
\left( \Psi^{(0)}_{\nu,\sigma} \left( \mathbf{k}\right) \right)^{\dagger} \Psi^{(0)}_{\nu,\sigma'}\left( \mathbf{k}'\right) \simeq \\
 \simeq \chi_{\nu,\mathbf{k},\sigma}^{\dagger} \left( 1 - \frac{\hbar^{2} \left( \mathbf{k} - \mathbf{k}' \right)^{2} }{8m^{2}c^{2}} + i \frac{\hbar^{2}}{4m^{2}c^{2}} \left[ \mathbf{k} \times \mathbf{k}' \right] \bm{\hat{\sigma}} \right) \chi_{\nu,\mathbf{k}',\sigma'} , \; \nu = p,a , \\
\left( \Psi^{(0)}_{p,\sigma} \left( \mathbf{k}\right) \right)^{\dagger} \Psi^{(0)}_{a,\sigma'}\left( \mathbf{k}'\right) \simeq \frac{\hbar}{2mc} \chi_{p,\mathbf{k},\sigma}^{\dagger} \left( \mathbf{k} - \mathbf{k}' \right) \bm{\hat{\sigma}} \chi_{a, \mathbf{k}', \sigma'} , \\
\left( \Psi^{(0)}_{a,\sigma} \left( \mathbf{k}\right) \right)^{\dagger} \Psi^{(0)}_{p,\sigma'}\left( \mathbf{k}'\right) \simeq - \frac{\hbar}{2mc} \chi_{a,\mathbf{k},\sigma}^{\dagger} \left( \mathbf{k} - \mathbf{k}' \right) \bm{\hat{\sigma}} \chi_{p, \mathbf{k}', \sigma'} .
\end{array} 
\end{equation} 

Supposing that both ratios $ \mid V \mid /mc^{2} $ and $ \hbar k / mc $ are  of the same order $ \sim \lambda $, one can neglect renormalization of particles and antiparticles up to the second order, since, as it was indicated above (see Eqs. (\ref{W_p,a}) and (\ref{prbisp_nr})), the smallness of $ W^{(p)}_{\sigma ,\sigma'} \left( \mathbf{k} ,\mathbf{k}'\right) \sim \lambda^{3} $ exceeds this accuracy. In the result, the Hamiltonian of particles (\ref{tildeH_p}) in the non-relativistic approximation takes the form
%\footnote{
%For the sake of simplicity below we consider particles only, therefore, the index "$ p $" is omitted. The %Hamiltonian of antiparticles can be derived in a similar way. 
%}  
\begin{equation}
\label{H_nr} 
\mathrm{H} = \sum_{\mathbf{k},\sigma} \left( \frac{\hbar^{2} k^{2}}{2m} \left( 1 - \lambda_{SO}^{(0)} k^{2} \right) a_{\mathbf{k},\sigma}^{\dagger} a_{\mathbf{k},\sigma} + \frac{1}{L^{3}} \sum_{\mathbf{k}',\sigma'} \mathcal{V}_{\sigma ,\sigma'} \left(\mathbf{k} ,\mathbf{k}' \right) a_{\mathbf{k},\sigma}^{\dagger} a_{\mathbf{k}',\sigma'} \right) ,
\end{equation} 
where their energy is counted from the  rest energy $ mc^{2} $. The scattering amplitudes in the second sum, according to  (\ref{calV_p,a}) and (\ref{prbisp_nr}), are given by expression 
\begin{equation}
\label{calV_nr}  
\mathcal{V}_{\sigma ,\sigma'} \left(\mathbf{k} ,\mathbf{k}' \right) = V \left(\mathbf{k} - \mathbf{k}' \right) \chi_{\mathbf{k},\sigma}^{\dagger} \left( 1 - \frac{\lambda_{SO}^{(0)} }{2} \left( \mathbf{k} - \mathbf{k}' \right)^{2} + i  \lambda_{SO}^{(0)} \left[ \mathbf{k} \times \mathbf{k}' \right] \bm{\hat{\sigma}} \right) \chi_{\mathbf{k}',\sigma'} .
\end{equation} 
For the sake of simplicity here and below we consider particles only, therefore, the index "$ p $" is omitted. The Hamiltonian for antiparticles can be derived in a similar way. 

Operators $ a_{\mathbf{k},\sigma}^{\dagger} $ and $ a_{\mathbf{k},\sigma} $ in the Hamiltonian (\ref{H_nr}) are creation and annihilation operators of an electron with the momentum $ \hbar \mathbf{k} $ in the spin state corresponding to the invariant (\ref{op-Inv}). This means that the spinors $ \chi_{\sigma}\left( \mathbf{k} \right) $ in Eq. (\ref{calV_nr}) are the solutions of Eq. (\ref{egnspinor_p,a}) (with $ \nu = p $), hence, they are eigen spinors of the matrix 
$ \mathbf{r} \left( \mathbf{k} \right) \bm{\hat{\sigma}} $ with $ \mathbf{r} \left( \mathbf{k} \right) \equiv \mathbf{r}_p $ (see Eq. (\ref{r_+,-})). Therefore, in the non-relativistic limit, vector $ \mathbf{r} \left( \mathbf{k} \right) $ also has to be expanded with respect to the small parameter. In particular, up to the second order it is given by the expression 
\begin{equation}
\label{vecr_nr} 
\begin{array}{c}
\mathbf{r} \left( \mathbf{k} \right) \simeq \mathbf{r}^{(0)}_{\mathbf{k}} + 2 \lambda_{SO}^{(0)} \mathbf{r}^{(2)}_{\mathbf{k}}  ,  \\
\mathbf{r}^{(0)}_{\mathbf{k}} = \mathbf{r}_{\bm{\mu}} \left( \mathbf{k} \right) + \mathbf{r}_{\bm{\mathcal{S}}} \left( \mathbf{k} \right) , \quad \mathbf{r}^{(2)}_{\mathbf{k}} = \left( \mathbf{r}_{\bm{\mathcal{S}}} \left( \mathbf{k} \right) \mathbf{k} \right) \mathbf{k} - \left[ \mathbf{r}_{\bm{\mu}}\left( \mathbf{k} \right) \times \mathbf{k} \right] \times \mathbf{k} ,
\end{array}
\end{equation}
which shows that in such a case the spinors also contain the relativistic corrections.

As it has been pointed, the direction of the vector $ \mathbf{r}\left( \mathbf{k} \right) $ determines spin quantization axis for each $ \mathbf{k} $-state. This axis can be obtained from the initial $ z $-axis via the rotation using some operator  $ \hat{\omega} $. Then, the spin functions   $ \chi_{\sigma} $ transform into functions $ \tilde{\chi}_{\sigma} = \hat{\omega} \chi_{\sigma} $, and matrix $ \mathbf{r}\left( \mathbf{k} \right) \bm{\hat{\sigma}} $, transforms into matrix $ \hat{\omega} \mathbf{r}\left( \mathbf{k} \right) \bm{\hat{\sigma}} \hat{\omega}^{\dagger} = r\left( \mathbf{k} \right) \hat{\sigma}_{z} $. In the latter expression the matrix  $ \hat{\sigma}_{z} $ refers to a new (rotated  or local) coordinate system. Therefore, its spinors $ \tilde{\chi}_{\sigma} $ are eigen spinors for this Pauli matrix, and, hence, have the form $ \tilde{\chi}_{\uparrow} $ and $ \tilde{\chi}_{\downarrow} $, as it has been underlined above. From them the spinors in the initial coordinate system can be calculated by the action of the operator $\hat{\omega} \left( \mathbf{k} \right)$:  
\begin{equation}
\label{spinor_pl,m} 
\chi_{\sigma}\left( \mathbf{k} \right) = \hat{\omega}^{\dagger} \left( \mathbf{k} \right) \tilde{\chi}_{\sigma} , \quad 
\chi_{+}\left( \mathbf{k} \right)  = \hat{\omega}^{\dagger} \left( \mathbf{k} \right) \tilde{\chi}_{\uparrow}  , \quad \chi_{-}\left( \mathbf{k} \right)  = \hat{\omega}^{\dagger} \left( \mathbf{k} \right) \tilde{\chi}_{\downarrow} .
\end{equation}

It can be shown that the corresponding rotation (neglecting the common phase multiplier) can be realized by the above operator 
\begin{equation}
\label{omega_rot} 
\hat{\omega} \left( \mathbf{k} \right) = \sqrt{\frac{r\left( \mathbf{k} \right) + \mathbf{e}_{z} \mathbf{r}\left( \mathbf{k} \right) }{2r\left( \mathbf{k} \right)}} \left( \hat{I}_{2} + i \frac{ \mathbf{d}\left( \mathbf{k} \right) \bm{\hat{\sigma}} }{r\left( \mathbf{k} \right) + \mathbf{e}_{z} \mathbf{r}\left( \mathbf{k} \right)} \right) ,  \quad \mathbf{d}\left( \mathbf{k} \right) \equiv \left[ \mathbf{e}_{z} \times \mathbf{r}\left( \mathbf{k} \right) \right] .
\end{equation} 
Using vector $ \mathbf{r} $ in the form (\ref{vecr_nr}) and choosing without the loss of the generality $ \mathbf{r}^{(0)}_{\mathbf{k}} $ in (\ref{vecr_nr}) as a unit vector, one can write down the expansion for the matrix  (\ref{omega_rot}) with respect to the parameter $ \hbar k/mc $:
\[
\hat{\omega}\left( \mathbf{k} \right) \simeq \hat{\omega}^{(0)}\left( \mathbf{k} \right) + 2 \lambda_{SO}^{(0)} \hat{\omega}^{(2)}\left( \mathbf{k} \right) .
\]
where  
\begin{equation}
\label{omega_0} 
\hat{\omega}^{(0)}\left( \mathbf{k} \right) = \sqrt{\frac{1 + z^{(0)}_{\mathbf{k}} }{2}} \left( \hat{I}_{2} + i \frac{\mathbf{d}^{(0)}_{\mathbf{k}} \bm{\hat{\sigma}} }{1 + z^{(0)}_{\mathbf{k}}} \right)  ,\quad \left( \hat{\omega}^{(0)}\right)^{\dagger} \hat{\omega}^{(0)} = \hat{I}_{2} ,
\end{equation}
\begin{equation}
\label{omega_2} 
\hat{\omega}^{(2)}\left( \mathbf{k} \right) = \sqrt{\frac{1 + z^{(0)}_{\mathbf{k}} }{2}} \left( Q_{-} \hat{I}_{2} + i \frac{ \left[ \mathbf{d}^{(2)}_{\mathbf{k}} - Q_{+} \mathbf{d}^{(0)}_{\mathbf{k}} \right] \bm{\hat{\sigma}} }{1 + z^{(0)}_{\mathbf{k}}} \right)   , 
\end{equation} 
and, according to (\ref{vecr_nr}), the notations 
\begin{equation}
\label{notat} 
\begin{array}{c}
Q_{\pm} = \frac{1}{2} \left( \frac{r^{(2)}_{ \mathbf{k}} + z^{(2)}_{\mathbf{k}}}{r^{(0)}_{\mathbf{k}} + z^{(0)}_{\mathbf{k}}} \pm \frac{r^{(2)}_{ \mathbf{k}}}{r^{(0)}_{\mathbf{k}}} \right)  , \quad  r^{(2)}_{ \mathbf{k}} = \mathbf{r}^{(0)}_{\mathbf{k}} \mathbf{r}^{(2)}_{\mathbf{k}} ,\\
 \mathbf{d}^{(j)}_{\mathbf{k}} = \mathbf{e}_{z} \times \mathbf{r}^{(j)}_{ \mathbf{k}} , \quad z^{(j)}_{\mathbf{k}} = \mathbf{r}^{(j)}_{\mathbf{k}} \mathbf{e}_{z} , \; j = 1,2 \, 
\end{array}
\end{equation}
are used. Matrix $ \hat{\omega}\left( \mathbf{k} \right) $ is unitary up to the required accuracy due to the relation
\begin{equation}
\label{omeg_2xomeg1} 
\left( \hat{\omega}^{(2)}\left(\mathbf{k} \right)\right) ^{\dagger} \hat{\omega}^{(0)}\left(\mathbf{k} \right) = - \frac{i}{2} \bm{\Lambda} \left( \mathbf{k} \right) \bm{\hat{\sigma}} = - \left( \hat{\omega}^{(0)}\left(\mathbf{k} \right)\right) ^{\dagger} \hat{\omega}^{(2)}\left(\mathbf{k} \right) 
\end{equation}
in which
$$ 
\bm{\Lambda} \left( \mathbf{k} \right) = \mathbf{d}^{(2)}_{\mathbf{k}} - \frac{r^{(2)}_{\mathbf{k}} + z^{(2)}_{\mathbf{k}}}{1 + z^{(0)}_{\mathbf{k}}} \mathbf{d}^{(0)}_{\mathbf{k}} - \frac{\mathbf{d}^{(2)}_{\mathbf{k}} \times \mathbf{d}^{(0)}_{\mathbf{k}}}{1 + z^{(0)}_{\mathbf{k}}} .
$$
Using  equalities of vector algebra and definitions (\ref{notat}), this expression can be reduced to the following form: 
\begin{equation}
\label{D(k)} 
\bm{\Lambda} \left( \mathbf{k} \right) = \mathbf{r}^{(0)}_{\mathbf{k}} \times \mathbf{r}^{(2)}_{\mathbf{k}} + \frac{ \mathbf{r}^{(0)}_{\mathbf{k}} \left[ \mathbf{e}_{z} \times \mathbf{r}^{(2)}_{\mathbf{k}}\right]}{1 + z^{(0)}\left( \mathbf{k} \right)} \mathbf{r}^{(0)}_{\mathbf{k}} .
\end{equation}

To calculate the relativistic corrections to the spinors, we will use the expression (\ref{spinor_pl,m}) and matrix expansion (\ref{omega_rot}) taking into account that matrix (\ref{omega_0}) is unitary: 
\begin{equation}
\begin{array}{c}
\chi_{\mathbf{k},\sigma} = \left[ \left( \hat{\omega}^{(0)} \left( \mathbf{k} \right)\right)^{\dagger} + 2 \lambda_{SO}^{(0)} \left( \hat{\omega}^{(2)} \left( \mathbf{k} \right)\right) ^{\dagger} \right] \tilde{\chi}_{\sigma} = \\ 
= \left[ \left( \hat{\omega}^{(0)} \left( \mathbf{k} \right)\right)^{\dagger} + 2 \lambda_{SO}^{(0)} \left( \hat{\omega}^{(2)} \left( \mathbf{k} \right)\right) ^{\dagger} \hat{\omega}^{(0)}\left(\mathbf{k} \right) \left( \hat{\omega}^{(0)} \left( \mathbf{k} \right)\right) ^{\dagger} \right] \tilde{\chi}_{\sigma} = \\
= \left( \hat{I}_{2} - i \lambda_{SO}^{(0)} \bm{\Lambda} \left( \mathbf{k} \right) \bm{\hat{\sigma}} \right) \chi^{(0)}_{\mathbf{k},\sigma} , 
\end{array}
\label{spinor}
\end{equation}
where the equality (\ref{omeg_2xomeg1}) has been used and $ \chi^{(0)}_{\mathbf{k},\sigma} = \left( \hat{\omega}^{(0)} \left( \mathbf{k} \right)\right) ^{\dagger} \tilde{\chi}_{\sigma} $. 

After the substitution of obtained expression for $ \chi_{\mathbf{k},\sigma} $ into Eq. (\ref{calV_nr}), the scattering amplitudes become  
\begin{equation}
\label{calV_nrl2} 
\begin{array}{c}
\mathcal{V}_{\sigma ,\sigma'} \left(\mathbf{k} ,\mathbf{k}' \right) = V \left(\mathbf{k} - \mathbf{k}' \right) \left( \chi^{(0)}_{\mathbf{k},\sigma} \right)^{\dagger} \hat{\mathcal{V}} \left(\mathbf{k} ,\mathbf{k}' \right) \chi^{(0)}_{\mathbf{k}',\sigma'} , \\
\hat{\mathcal{V}} \left(\mathbf{k} ,\mathbf{k}' \right) = 
\left( 1 - \frac{\lambda_{SO}^{(0)}  }{2} \left( \mathbf{k} - \mathbf{k}' \right)^{2} \right)\hat{I}_{2} + i \lambda_{SO}^{(0)} \bm{\Lambda} \left( \mathbf{k},\mathbf{k}' \right) \bm{\hat{\sigma}} ,  \\
\bm{\Lambda} \left( \mathbf{k},\mathbf{k}' \right) = \bm{\Lambda}_{Th} \left( \mathbf{k},\mathbf{k}' \right) + \bm{\Lambda}_{BEL} \left( \mathbf{k},\mathbf{k}' \right) , \\
\bm{\Lambda}_{Th} \left( \mathbf{k},\mathbf{k}' \right) = \left[ \mathbf{k} \times \mathbf{k}' \right] , \quad \bm{\Lambda}_{BEL} \left( \mathbf{k},\mathbf{k}' \right) = \bm{\Lambda} \left( \mathbf{k} \right) - \bm{\Lambda} \left( \mathbf{k}' \right)  .
\end{array}
\end{equation} 

The Hamiltonian  (\ref{H_nr}) with matrix elements $ \mathcal{V}_{\sigma ,\sigma'} \left(\mathbf{k} ,\mathbf{k}' \right) $ in the form (\ref{calV_nrl2}) contains {\it{all}} relativistic corrections up to the second order both in the kinetic and potential energies. Among them there are the well known corrections, such as Darwin correction (the second term at the unit matrix), and Thomas correction $\sim  
\bm{\Lambda}_{Th} \left( \mathbf{k},\mathbf{k}' \right)$ \cite{Bete,RelQuant,Davydov,Messia}. Recall, the term which contains the spin operator $ \bm{\hat{\sigma}} $ in (\ref{calV_nrl2}) is called SOI, since it brings the interdependence between particle's spin and motion in a inhomogeneous potential. One can see, however, this interaction includes not only Thomas correction, but also one more term $\sim \bm{\Lambda}_{BEL} \left( \mathbf{k},\mathbf{k}' \right) $ which is fully determined by the relativistic corrections to the spinors. So far to our knowledge, the latter correction has not been derived before and its role has not been investigated.  
 
Comparing all relativistic corrections to the Schr{\"o}dinger Hamiltonian, it is clear that in the case of the non-relativistic motion the correction to the kinetic energy is negligibly small. Besides, the second order is, strictly speaking, satisfactory only under the condition of small changes of the fields on the distances of the order of the Compton wavelength \cite{RelQuant}. This condition, as a rule, is fulfilled for particles in macroscopic (including crystal) fields, hence, the Darwin correction in the potential energy is also small and usually (as in Eq. (\ref{H_nrl}) is omitted. This explains why the Schr{\"o}dinger operator is adoptedly corrected with the one SOI term, only. In particular, SOI provides relatively small, but experimentally observable spin splitting of energy levels (bands) and, essentially, determines spin quantization axis which is not arbitrary, but depends on the form of the potential. Just the spin polarization of particles in condensed systems is now the subject of numerous studies and this is why investigation of  SOI effects is important.   

To proceed, let us rewrite the Hamiltonian (\ref{H_nr}) in the following form: 
\begin{equation}
\label{H_nrl2} 
\mathrm{H} = \sum_{\mathbf{k},\sigma} \left( \frac{\hbar^{2} k^{2}}{2m} a_{\mathbf{k},\sigma}^{\dagger} a_{\mathbf{k},\sigma} + \frac{1}{L^{3}} \sum_{\mathbf{k}',\sigma'} V \left(\mathbf{k} - \mathbf{k}' \right) \mathcal{V}_{\sigma ,\sigma'} \left(\mathbf{k} ,\mathbf{k}' \right) a_{\mathbf{k},\sigma}^{\dagger} a_{\mathbf{k}',\sigma'} \right)  ,
\end{equation}  
where $\hat{\sigma }$-dependent quantity  
\begin{equation}
\label{opcalV_nr} 
\mathcal{V}_{\sigma ,\sigma'} \left(\mathbf{k} ,\mathbf{k}' \right) =  \left( \chi^{(0)}_{\mathbf{k},\sigma} \right)^{\dagger}  \hat{\mathcal{V}} \left(\mathbf{k} ,\mathbf{k}' \right)  \chi^{(0)}_{\mathbf{k}',\sigma '} \approx   \left( \chi^{(0)}_{\mathbf{k},\sigma} \right)^{\dagger} \left( \hat{I}_{2} + i \lambda_{SO}^{(0)} \bm{\Lambda} \left( \mathbf{k},\mathbf{k}' \right) \bm{\hat{\sigma}} \right) \chi^{(0)}_{\mathbf{k}',\sigma'}, 
\end{equation}
includes the matrix of the spin-orbit scattering. The latter contains both corrections, according to the definitions (\ref{calV_nrl2}): $ \bm{\Lambda}_{Th} \left( \mathbf{k},\mathbf{k}' \right) $ and $ \bm{\Lambda}_{BEL} \left( \mathbf{k},\mathbf{k}' \right) $. In (\ref{opcalV_nr}) spinor $ \chi^{(0)}_{\mathbf{k},\sigma}  $, as in Eq. (\ref{spinor}), is a non-relativistic eigen spinor of the matrix $ \mathbf{r}^{(0)}_{\mathbf{k}} \bm{\hat{\sigma}} $. Hence, the unit vector $ \mathbf{r}^{(0)}_{\mathbf{k}} = \gamma_{x}\left( \mathbf{k} \right) \mathbf{e}_{x} + \gamma_{y}\left( \mathbf{k} \right) \mathbf{e}_{y} + \gamma_{z}\left( \mathbf{k} \right) \mathbf{e}_{z} $ determines the spin quantization axis with the guiding cosines $ \gamma_{j}\left( \mathbf{k} \right) $ that are given by the sum of the corresponding vector coefficients $ \mathbf{r}_{\bm{\mu}}(\mathbf{k}) $ and $ \mathbf{r}_{\bm{\mathcal{S}}}(\mathbf{k}) $ from Eq. (\ref{Inv_g2}). It is easy to find that 
\[
\chi_{\mathbf{k},\sigma}^{(0)}\left(\varphi_{\mathbf{k}},  \vartheta_{\mathbf{k}}  \right) \equiv  \chi_{\mathbf{k},\sigma}^{(0)} = e^{i\sigma \varphi_{\mathbf{k}}/2} 
\left( \begin{array}{c}
\sigma \sqrt{\frac{1+\sigma \gamma_{z}\left( \mathbf{k} \right)}{2}} e^{-i\varphi_{\mathbf{k}}/2} \\ 
\sqrt{\frac{1-\sigma \gamma_{z}\left( \mathbf{k} \right)}{2}} e^{i\varphi_{\mathbf{k}}/2}
\end{array} \right) , \quad \sigma = \pm 1 , 
\]
where $ \tan \varphi_{\mathbf{k}}  = \gamma_{y}\left( \mathbf{k} \right)/\gamma_{x}\left( \mathbf{k} \right) $.  Using the equalities $ \gamma_{x}\left( \mathbf{k} \right) = \sin \vartheta_{\mathbf{k}} \cos \varphi_{\mathbf{k}} $, $ \gamma_{y}\left( \mathbf{k} \right) = \sin \vartheta_{\mathbf{k}} \sin \varphi_{\mathbf{k}} $, $ \gamma_{z}\left( \mathbf{k} \right) = \cos \vartheta_{\mathbf{k}} $, one can introduce spin variables (angles) $ \vartheta_{\mathbf{k}} $ and $ \varphi_{\mathbf{k}} $, on which the spinors in the zero approximation depend: 
\begin{equation}
\label{chi(0)} 
\chi_{\mathbf{k},+}^{(0)} =
\left( \begin{array}{c} 
\cos \frac{\vartheta_{\mathbf{k}}}{2} \\ 
e^{i\varphi_{\mathbf{k}} } \sin \frac{\vartheta_{\mathbf{k}}}{2}
\end{array} \right) , \quad \chi_{\mathbf{k},-}^{(0)} =
\left( \begin{array}{c} 
-e^{-i\varphi_{\mathbf{k}} } \sin \frac{\vartheta_{\mathbf{k}}}{2} \\ \cos \frac{\vartheta_{\mathbf{k}}}{2}
\end{array} \right)  .
\end{equation} 
In fact, these variables can be considered and indeed are the free parameters. 

Such a picture is radically changed in the presence of the external potential when these parameters become fixed. This follows from the fact that the stationary electron states can be found diagonalizing the Hamiltonian (\ref{H_nrl2}) to the form 
\begin{equation}
\label{H_diag} 
\mathrm{H} = \sum_{\nu} E_{\nu} a_{\nu}^{\dagger} a_{\nu} 
\end{equation}
where $ E_{\nu} $ are the energies of these states, and index $ \nu = n,\sigma $ is a set of the quantum numbers with the spin number $ \sigma $ taking two values.  

For diagonalization of the Hamiltonian (\ref{H_nrl2}) it is convenient to choose the spinors (\ref{chi(0)}) for which the matrix element (\ref{opcalV_nr}) is proportional to $ \delta_{\sigma ,\sigma'} $. They diagonalize the invariant (\ref{op-Inv}), $ \mathfrak{I}_{gen}^{(p)} $, and in non-relativistic limit are determined as eigen spinors of the operator $ \mathbf{r}^{(0)}_{\mathbf{k}} \bm{\hat{\sigma}} $ with $ \mathbf{r}^{(0)}_{\mathbf{k}}$ given in Eq. (\ref{vecr_nr}). Therefore the operator $ \mathfrak{I}_{gen}^{(p)} $ should commute with the Hamiltonian (\ref{H_nrl2}). This condition implies the restrictions on the free parameters values, or the vectors $ \mathbf{r}_{\bm{\mu}} (\mathbf{k})$ and $ \mathbf{r}_{\bm{\mathcal{S}}} (\mathbf{k})$; on the other  side, such a condition results in the equation which determines the vector $ \mathbf{r}^{(0)}_{\mathbf{k}} $ that coincides with the spin quantization axis and, hence, the spin variable dependence appears. 

The analysis of obtained equation depends on the structure of the Fourier transformation $ V \left(\mathbf{k} - \mathbf{k}' \right) $ of the potential and has to be performed in the general case in the curvilinear coordinate system over the surface in which the equipotential surfaces are formed, and perpendicular to them directions are determined by the potential gradients in each point. This allows to find the solutions at least in the vicinity of singular points or lines that are characteristic for this  potential, if not in their whole definition area. The main difficulty here is the dependence of the local reper orientation on the coordinates of the studied spatial point. This is the reason why it is impossible to find the general expression for SOI in the coordinate space for the general form potential. Below we consider one of the simplest but nevertheless actual case of electrons in quasi-2D system. 

%\section{Two-dimensional motion \label{2D}}
\section{6. Two-dimensional motion}

Any 2D system, in fact is  \textit{quasi}-2D  with a finite spatial width, for example, in  $ z $ direction.  In such a case the potential in the Hamiltonian (\ref{H_nrl2}) reflects a translational symmetry in $xy$-plane: $ V\left( \mathbf{r}\right)= V\left( \mathbf{r_{\perp}}, z \right)= V\left( \mathbf{r_{\perp}}+ \mathbf{l_{\perp}}, z \right)$ where $ \mathbf{l_{\perp}}=\sum_{j=1,2}l_j \mathbf{a}_j $ is a translation vector in 2D lattice structure with basis vectors $ \mathbf{a}_j $ ($l_j$  are integer numbers). Such potentials can be represented by Fourier series 
\begin{equation}
\label{V_Four} 
V\left( \mathbf{r}_{\perp},z \right) = \sum_{\mathbf{g}_{\perp}} V_{\mathbf{g}_{\perp}}(z) e^{i\mathbf{g}_{\perp} \mathbf{r}_{\perp}} = V_{0}(z) + V_{\perp}(\mathbf{r}_{\perp},z) , \quad V_{\perp}(\mathbf{r}_{\perp},z) = \sum_{\mathbf{g}_{\perp}\neq 0} V_{\mathbf{g}_{\perp}}(z) e^{i\mathbf{g}_{\perp} \mathbf{r}_{\perp}}
\end{equation}
where vectors $ \mathbf{g}_{\perp} = \sum_{j=1,2} n_{j} \mathbf{b}_{j} $ are defined by basis vectors $ \mathbf{b}_{j} $ of the corresponding reciprocal lattice with $ n_j $ being integer numbers. Expansion coefficients in Eq. (\ref{V_Four}), which in a general case can depend on $z$, are determined by the formula
\[
V_{\mathbf{g}_{\perp}}(z) = \frac{1}{S_{\textrm {cell}}} \int V \left( \mathbf{r}_{\perp},z\right) e^{-i\mathbf{g}_{\perp} \mathbf{r}_{\perp}} dx dy 
\]
where integration is carried out over unit cell of an area $ S_{\textrm {cell}} $. 

According to (\ref{V_Four}), the potential can be represented as a sum of two components, namely zero harmonic $ V_{0}(z) $ and periodic in plane part $ V_{\perp}(\mathbf{r}_{\perp},z) $. In this case the Fourier transformation of the potential is 
\[
 V\left(\mathbf{k} - \mathbf{k}' \right) = L^{2} \delta_{\mathbf{k}_{\perp},\mathbf{k}_{\perp}'} V_0\left( k_{z} - k_{z}' \right) + V_{\perp}\left(\mathbf{k}_{\perp} - \mathbf{k}_{\perp}', k_{z} - k_{z}' \right) ,
\]
which allows to present the Hamiltonian (\ref{H_nrl2}) in the form $ \mathrm{H} = \mathrm{H}_{2D}^{(0)} + \mathrm{V}_{per} $ where 
\begin{equation}
\label{H_2D_tot} 
\begin{array}{c}
\mathrm{H}_{2D}^{(0)} = \sum_{\mathbf{k},\sigma} \left( \frac{\hbar^{2} \mathbf{k}^{2} }{2m} a_{\mathbf{k},\sigma}^{\dagger} a_{\mathbf{k},\sigma} + \frac{1}{L} \sum_{\mathbf{k}',\sigma'} \delta_{\mathbf{k}_{\perp},\mathbf{k}_{\perp}'} V_{0} \left(k_{z} - k_{z}' \right) \mathcal{V}_{\sigma ,\sigma'} \left(\mathbf{k},\mathbf{k}' \right) a_{\mathbf{k},\sigma}^{\dagger} a_{\mathbf{k}',\sigma'} \right) , \\
\mathrm{V}_{per} = \frac{1}{L^{3}} \sum_{\mathbf{k},\mathbf{k}'} \sum_{\sigma ,\sigma'} V_{\perp} \left(\mathbf{k} - \mathbf{k}' \right) \mathcal{V}_{\sigma ,\sigma'} \left(\mathbf{k} ,\mathbf{k}' \right) a_{\mathbf{k},\sigma}^{\dagger} a_{\mathbf{k}',\sigma'} .  
\end{array}
\end{equation}
with $ \mathcal{V}_{\sigma ,\sigma'} \left(\mathbf{k},\mathbf{k}' \right) $ is given in Eq. (\ref{opcalV_nr}) and, as above,  $ \mathbf{k} = \mathbf{k}_{\perp} + k_{z}\mathbf{e}_{z} $. 

As the first stage of finding the energy spectrum let us consider the Hamiltonian $ \mathrm{H}_{2D}^{(0)} $ only. The matrix element (\ref{opcalV_nr}) in $ \mathrm{H}_{2D}^{(0)} $ is calculated at $ \mathbf{k} = \mathbf{k}_{\perp} + k_{z} \mathbf{e}_{z} $ and $ \mathbf{k}' = \mathbf{k}_{\perp} + k_{z}' \mathbf{e}_{z} $. This leads to the following expressions for Thomas part $ \bm{\Lambda}_{Th} \left( \mathbf{k}, \mathbf{k}' \right) = \left( k_{z} - k_{z}' \right) \left[ \mathbf{e}_{z} \times \mathbf{k}_{\perp} \right] $ and additional part $ \bm{\Lambda}_{BEL} \left( \mathbf{k},\mathbf{k}' \right) = \bm{\Lambda} \left( \mathbf{k}_{\perp},k_{z} \right) - \bm{\Lambda} \left( \mathbf{k}_{\perp},k_{z}' \right) $ of SOI (see (\ref{calV_nrl2})). 

Spin states can be chosen arbitrary, and it is convenient to take such spinors $ \chi^{(0)}_{\mathbf{k},\sigma} $  (\ref{chi(0)}) that satisfy the condition $ \mathcal{V}_{\sigma ,\sigma'} \left(\mathbf{k}_{\perp},k_{z},k_{z}' \right) = \delta_{\sigma ,\sigma'} \mathcal{V}_{\sigma } \left(\mathbf{k}_{\perp},k_{z},k_{z}' \right) $. In view of the  matrix structure 
%in (\ref{opcalV_nr}) 
and ortho-normalization of spinors $ \chi^{(0)}_{\mathbf{k},\sigma} $, this condition is fulfilled, provided, first, the zero order spinors in (\ref{H_2D_tot}) does not depend on $ k_{z} $, that means the dependence  of its spin quantization axis on $ \mathbf{k}_{\perp} $ only. Secondly, diagonalization of $ \mathrm{H}_{2D}^{(0)} $ with respect to spin numbers is performed if the spinors  $ \chi^{(0)}_{\mathbf{k}_{\perp},\sigma} $ are the eigen spinors of the matrix $ \bm{\Lambda} \left( \mathbf{k},\mathbf{k}' \right) \bm{\hat{\sigma}} $. Since  $ \chi^{(0)}_{\mathbf{k}_{\perp},\sigma} $ are defined as eigen spinors of the matrix $ \mathbf{r}^{(0)}_{\mathbf{k}_{\perp}} \bm{\hat{\sigma}} $, this condition is fulfilled provided the commutator $ \left[ \bm{\Lambda} \left( \mathbf{k},\mathbf{k}' \right) \bm{\hat{\sigma}} , \mathbf{r}^{(0)}_{\mathbf{k}_{\perp}} \bm{\hat{\sigma}} \right] $ is equal zero which leads to the equality $ \bm{\Lambda} \left( \mathbf{k},\mathbf{k}' \right) \times \mathbf{r}^{(0)}_{\mathbf{k}_{\perp}} = 0 $.

Taking into account Eq. (\ref{D(k)}), Eq. (\ref{vecr_nr}) and condition $ \mathbf{r}^{(0)}_{\mathbf{k}} = \mathbf{r}^{(0)}_{\mathbf{k}_{\perp}} $ the following expressions for the additional correction  can be obtained 
\begin{equation}
\label{D_EBL-2D} 
\bm{\Lambda}_{BEL} \left( \mathbf{k},\mathbf{k}' \right) = \mathbf{r}^{(0)}_{\mathbf{k}_{\perp}} \times \bm{\lambda} \left( \mathbf{k}_{\perp}, k_{z},k_{z}' \right) + \frac{ \mathbf{r}^{(0)}_{\mathbf{k}_{\perp}} \left[ \mathbf{e}_{z} \times \bm{\lambda} \left( \mathbf{k}_{\perp}, k_{z},k_{z}' \right) \right]}{1 + z^{(0)}_{\mathbf{k}_{\perp}}} \mathbf{r}^{(0)}_{\mathbf{k}_{\perp}} ,
\end{equation}
where 
\begin{equation}
\label{lambda_2D} 
\begin{array}{c}
\bm{\lambda} \left( \mathbf{k}_{\perp}, k_{z},k_{z}' \right) = \mathbf{r}^{(2)}_{\mathbf{k}_{\perp},k_{z}} - \mathbf{r}^{(2)}_{\mathbf{k}_{\perp},k_{z}'} = \left( k_{z} - k_{z}' \right) \bm{\lambda} \left( \mathbf{k}_{\perp} \right) + \left( k_{z}^{2} - \left. k_{z}'\right.^{2} \right) \bm{\lambda}' \left( \mathbf{k}_{\perp} \right) , \\ 
\bm{\lambda} \left( \mathbf{k}_{\perp} \right) = \left[ \mathbf{r}_{\bm{\mathcal{S}}}\left( \mathbf{k}_{\perp} \right) \mathbf{e}_{z} - \mathbf{r}_{\bm{\mu}}\left( \mathbf{k}_{\perp} \right)  \mathbf{e}_{z} \right] \mathbf{k}_{\perp} + \left[ \mathbf{r}_{\bm{\mathcal{S}}}\left( \mathbf{k}_{\perp} \right) \mathbf{k}_{\perp} - \mathbf{r}_{\bm{\mu}}\left( \mathbf{k}_{\perp} \right) \mathbf{k}_{\perp} \right]  \mathbf{e}_{z} , \\
\bm{\lambda}' \left( \mathbf{k}_{\perp} \right) = \left( \mathbf{r}_{\bm{\mathcal{S}}}\left( \mathbf{k}_{\perp} \right) \mathbf{e}_{z} \right) \mathbf{e}_{z} - \left[ \mathbf{r}_{\bm{\mu}}\left( \mathbf{k}_{\perp} \right) \times \mathbf{e}_{z} \right] \times \mathbf{e}_{z} . 
\end{array}
\end{equation} 
Therefore, in this 2D case the SOI vector (see (\ref{calV_nrl2})) in Eq. (\ref{opcalV_nr}) is 
\[
\begin{array}{c}
\bm{\Lambda} \left( \mathbf{k},\mathbf{k}' \right)  = \\
= \left( k_{z} - k_{z}' \right) \left( \mathbf{e}_{z} \times \mathbf{k}_{\perp} + \mathbf{r}^{(0)}_{\mathbf{k}_{\perp}} \times \bm{\lambda} \left( \mathbf{k}_{\perp} \right) + \frac{ \mathbf{r}^{(0)}_{\mathbf{k}_{\perp}} \left[ \mathbf{e}_{z} \times \bm{\lambda} \left( \mathbf{k}_{\perp} \right) \right]}{1 + z^{(0)}_{\mathbf{k}_{\perp}}} \mathbf{r}^{(0)}_{\mathbf{k}_{\perp}} \right) + \\
+ \left( k_{z}^{2} - \left. k_{z}'\right.^{2} \right) \left( \mathbf{r}^{(0)}_{\mathbf{k}_{\perp}} \times \bm{\lambda}' \left( \mathbf{k}_{\perp} \right) + \frac{ \mathbf{r}^{(0)}_{\mathbf{k}_{\perp}} \left[ \mathbf{e}_{z} \times \bm{\lambda}' \left( \mathbf{k}_{\perp} \right) \right]}{1 + z^{(0)}_{\mathbf{k}_{\perp}}} \mathbf{r}^{(0)}_{\mathbf{k}_{\perp}} \right) 
\end{array}
\]
from which the condition 
\begin{equation}
\label{cond2D} 
\begin{array}{c}
\bm{\Lambda} \left( \mathbf{k},\mathbf{k}' \right) \times \mathbf{r}^{(0)}_{\mathbf{k}_{\perp}} = \left( k_{z} - k_{z}' \right) \left[ \left[ \mathbf{e}_{z} \times \mathbf{k}_{\perp} \right] \times \mathbf{r}^{(0)}_{\mathbf{k}_{\perp}} + \bm{\lambda} \left( \mathbf{k}_{\perp} \right) - \left( \bm{\lambda} \left( \mathbf{k}_{\perp} \right) \mathbf{r}^{(0)}_{\mathbf{k}_{\perp} } \right) \mathbf{r}^{(0)}_{\mathbf{k}_{\perp}} \right]  + \\
+ \left( k_{z}^{2} - \left. k_{z}'\right.^{2} \right) \left[  \bm{\lambda}' \left( \mathbf{k}_{\perp} \right) - \left( \bm{\lambda}' \left( \mathbf{k}_{\perp} \right) \mathbf{r}^{(0)}_{\mathbf{k}_{\perp}} \right) \mathbf{r}^{(0)}_{\mathbf{k}_{\perp}} \right] = 0 
\end{array}
\end{equation}
is obtained. It can be satisfied only when the equality $ \bm{\lambda}'\left( \mathbf{k}_{\perp} \right) = 0 $ is valid. The latter, according to Eq. (\ref{lambda_2D}), implies constrains on the vectors $ \mathbf{r}_{\bm{\mu}}\left( \mathbf{k}_{\perp} \right) $ and $ \mathbf{r}_{\bm{\mathcal{S}}}\left( \mathbf{k}_{\perp} \right) $, which have to satisfy the equations
\[
\mathbf{r}_{\bm{\mu}}\left( \mathbf{k}_{\perp} \right) \times \mathbf{e}_{z} = 0 , \quad 
\mathbf{r}_{\bm{\mathcal{S}}}\left( \mathbf{k}_{\perp} \right) \mathbf{e}_{z} = 0 .
\]
This immediately  leads to expressions $ \mathbf{r}_{\bm{\mathcal{S}}}\left( \mathbf{k}_{\perp} \right) = x_{\bm{\mathcal{S}}}\left( \mathbf{k}_{\perp} \right) \mathbf{e}_{x} + y_{\bm{\mathcal{S}}}\left( \mathbf{k}_{\perp} \right) \mathbf{e}_{y} $ and $ \mathbf{r}_{\bm{\mu}}\left( \mathbf{k}_{\perp} \right) = z_{\bm{\mu}}\left( \mathbf{k}_{\perp} \right) \mathbf{e}_{z} $. Hence, the guiding cosines of a unit vector (cp. Eq. (\ref{vecr_nr})) 
\begin{equation}
\label{r_mu,r_S_2D} 
\mathbf{r}^{(0)}_{ \mathbf{k}_{\perp}} = \mathbf{r}_{\bm{\mu}}\left( \mathbf{k}_{\perp} \right) + \mathbf{r}_{\bm{\mathcal{S}}}\left( \mathbf{k}_{\perp} \right) = x_{\bm{\mathcal{S}}}\left( \mathbf{k}_{\perp} \right) \mathbf{e}_{x} + y_{\bm{\mathcal{S}}} \left( \mathbf{k}_{\perp} \right) \mathbf{e}_{y} + z_{\bm{\mu}}\left( \mathbf{k}_{\perp} \right) \mathbf{e}_{z} 
\end{equation}
are determined by the three parameters $ \gamma_{x}\left( \mathbf{k}_{\perp} \right) = x_{\bm{\mathcal{S}}}\left( \mathbf{k}_{\perp} \right) $, $ \gamma_{y}\left( \mathbf{k}_{\perp} \right) = y_{\bm{\mathcal{S}}}\left( \mathbf{k}_{\perp} \right) $ and $ \gamma_{z}\left( \mathbf{k}_{\perp} \right) = z_{\bm{\mu}}\left( \mathbf{k}_{\perp} \right) $. With such allowed vectors values the term proportional to $ \left( k_{z} - k_{z}' \right) $ in relation (\ref{cond2D}), is identically equal zero. Therefore, there are no additional conditions for the parameters $ x_{\bm{\mathcal{S}}}\left( \mathbf{k}_{\perp} \right) $, $ y_{\bm{\mathcal{S}}}\left( \mathbf{k}_{\perp} \right) $, $ z_{\bm{\mu}}\left( \mathbf{k}_{\perp} \right) $, so that they are indeed arbitrary. 

Substituting the found values of vectors $ \mathbf{r}_{\bm{\mu}}\left( \mathbf{k}_{\perp} \right) $ and $ \mathbf{r}_{\bm{\mathcal{S}}}\left( \mathbf{k}_{\perp} \right) $ into (\ref{D_EBL-2D}), one calculates the  vector $ \bm{\Lambda} \left( \mathbf{k},\mathbf{k}' \right) $ which characterizes SOI in matrix (\ref{opcalV_nr}) for 2D case:
\begin{equation}
\label{F(k)} 
\bm{\Lambda}_{2D} \left( \mathbf{k} ,\mathbf{k}' \right) = \left( k_{z} - k_{z}' \right) f \left( \mathbf{k}_{\perp} \right) \mathbf{r}^{(0)}_{ \mathbf{k}_{\perp}} , \quad  f \left( \mathbf{k}_{\perp} \right) = \frac{ \left[ \mathbf{e}_{z} \times \mathbf{k}_{\perp} \right] \mathbf{r}^{(0)}_{ \mathbf{k}_{\perp}} }{1 + \mathbf{e}_{z} \mathbf{r}^{(0)} _{ \mathbf{k}_{\perp}}} .
\end{equation}
As seen, it provides the diagonal form of $ \mathcal{V}_{\sigma ,\sigma'} \left(\mathbf{k}_{\perp},k_{z},k_{z}' \right) $ with respect to $ \sigma $ and each spin state is described by its own Hamiltonian  
\[
\begin{array}{c}
\mathrm{H}_{\sigma} = \sum_{\mathbf{k}_{\perp}, k_{z}} \left( \frac{\hbar^{2} \left(  k_{z}^{2} + \mathbf{k}_{\perp}^{2} \right)}{2m} a_{\mathbf{k}_{\perp},k_{z},\sigma}^{\dagger} a_{\mathbf{k}_{\perp},k_{z},\sigma} + \right. \\
\left. + \frac{1}{L} \sum_{k_{z}'} V_{0} \left( k_{z} - k_{z}' \right) \left[  1 - i \sigma \left( k_{z} - k_{z}' \right) f \left( \mathbf{k}_{\perp} \right) \right]  a_{\mathbf{k}_{\perp},k_{z},\sigma}^{\dagger} a_{\mathbf{k}_{\perp},k_{z}',\sigma} \right) ,  
\end{array}
\]
so that $ \mathrm{H}_{2D}^{(0)} = \sum_{\sigma} \mathrm{H}_{\sigma} $. 

The further diagonalization of the Hamiltonian with respect to $ k_{z} $ projections can be performed using the unitary transformations 
\begin{equation}
\label{transf1_2D} 
a_{\mathbf{k}_{\perp},k_{z},\sigma} = \sum_{\nu} \psi_{\nu ,\mathbf{k}_{\perp}, \sigma}\left( k_{z} \right) a_{\nu ,\mathbf{k}_{\perp}, \sigma } , \quad \sum_{k_{z}}  \psi_{\nu ,\mathbf{k}_{\perp}, \sigma}^{\ast}\left( k_{z} \right) \psi_{\mathbf{k}_{\perp}, \nu' ,\sigma}\left( k_{z} \right) = \delta_{\nu,\nu'} .
\end{equation}
Here the coefficients $ \psi_{\nu ,\mathbf{k}_{\perp}, \sigma}\left( k_{z} \right) $ satisfy the equation 
\[
\frac{\hbar^{2} \left( \mathbf{k}_{\perp}^{2} + k_{z}^{2} \right)}{2m} \psi_{\nu ,\mathbf{k}_{\perp}, \sigma}\left( k_{z} \right) + 
\]
\[
+ \frac{1}{L} \sum_{k_{z}'} V_{0} \left( k_{z} - k_{z}' \right) \left( 1 - i \left( k_{z} - k_{z}' \right) \sigma \lambda_{SO}^{(0)} f \left( \mathbf{k}_{\perp} \right) \right)  \psi_{\nu ,\mathbf{k}_{\perp}, \sigma}\left( k_{z}' \right) = E_{\nu ,\sigma} \left( \mathbf{k}_{\perp} \right) \psi_{\nu ,\mathbf{k}_{\perp}, \sigma}\left( k_{z} \right) .
\]
Its solution can be found after the transition to the coordinate representation 
\[
\psi_{\nu,\mathbf{k}_{\perp} ,\sigma} (z) = \frac{1}{\sqrt{L}} \sum_{k_{z}} e^{ik_{z} z} \psi_{\nu,\mathbf{k}_{\perp} ,\sigma} (k_{z})
\]
which leads to the stationary 1D SchE for each spin state 
\begin{equation}
\label{1DSEq} 
\left(  - \frac{\hbar^{2}}{2m} \frac{d^{2}}{dz^{2}} + \frac{\hbar^{2} \mathbf{k}_{\perp}^{2} }{2m} + V_{0}\left( z \right) + \sigma \lambda_{SO}^{(0)} f \left( \mathbf{k}_{\perp} \right) \frac{dV_{0}(z)}{dz} \right) \psi_{\mathbf{k}_{\perp}, \sigma}\left( z \right) = E \psi_{\mathbf{k}_{\perp}, \sigma}\left( z \right) .
\end{equation}
The fourth term in the l.h.s. of the latter equation is appeared due to the SOI. Namely this non-relativistic equation determines the wave functions $ \psi_{\nu ,\mathbf{k}_{\perp}, \sigma}\left( z \right) $ and eigen values $ E = E_{\nu ,\sigma} \left( \mathbf{k}_{\perp} \right) $ with account of SOI and coincides with equation obtained in \cite{AoP2,FNT} where the general analytical solution of the DE was found for the given problem. Notice, there can be situations when  the electron spinor components will be represented as linear combination of functions $ \psi_{\nu ,\mathbf{k}_{\perp}, +}\left( z \right) $ and $ \psi_{\nu ,\mathbf{k}_{\perp}, -}\left( z \right) $.

In the case of the QW potential the discrete eigen numbers $ \nu = n $ correspond to the bound states of electrons which are trapped by a QW and propagate as free particles in its plane.  In fact, the ensemble of such electrons is a 2D electron gas. Obviously, the solutions of Eq.  (\ref{1DSEq}) depend on the Qw form. The presence  ($ V_{0}\left( -z \right) = V_{0}\left( z \right) $) or absence ($ V_{0}\left( -z \right) \neq V_{0}\left( z \right) $) of the inverse symmetry of the QW  is the main factor of Rashba spin splitting of 2D electron bands \cite{Fabian,Rashba,Winkler,Rashba16}: 
\begin{equation}
\label{E(k)_2D} 
E_{n ,\sigma} \left( \mathbf{k}_{\perp} \right) = E_{n} \left( 0 \right) + \frac{\hbar^{2} \mathbf{k}_{\perp}^{2} }{2m} - \sigma \lambda_{SO} f \left( \mathbf{k}_{\perp} \right) , 
 \quad \lambda_{SO} = \lambda_{SO}^{(0)} a_{QW}  .
\end{equation} 
Here function $ f \left( \mathbf{k}_{\perp} \right) $ (see (\ref{F(k)})) introduces the explicit dependence of the splitting on electron spin state $ \sigma = \pm 1 $.  Parameter $ a_{QW} $ in Eq. (\ref{E(k)_2D}) depends on the form of the QW \cite{AoP1,AoP2,FNT} and characterizes its asymmetry, which can be intrinsic or caused by external electric field perpendicular to  $ xy $-plane.  The expression  (\ref{E(k)_2D}) without any assumption about the connection with a spin state, is the basic for study of SOI effects in a 2D electron gas using the Hamiltonian of free 2D particles,
\begin{equation}
\label{H_2D_0} 
\mathrm{H}_{2D}^{(0)} = \sum_{n,\mathbf{k}_{\perp},\sigma} E_{n,\sigma} \left( \mathbf{k}_{\perp} \right) a_{n,\mathbf{k}_{\perp},\sigma}^{\dagger} a_{n,\mathbf{k}_{\perp},\sigma} .  
\end{equation} 
Here the operators $ a_{n,\mathbf{k}_{\perp},\sigma}^{\dagger} $ and $ a_{n,\mathbf{k}_{\perp},\sigma} $ are creation and annihilation operators of electrons with wave vector $ \mathbf{k}_{\perp} $ in spin states determined by the spinors  (\ref{chi(0)})  whose quantization axis is parallel to vector  $ \mathbf{r}^{(0)} _{ \mathbf{k}_{\perp}} $.

Usually the Hamiltonian of a 2D electron gas is written using the operators  $ a_{\mathbf{k}_{\perp},\uparrow} $ and $ a_{\mathbf{k}_{\perp},\downarrow} $ which are related to the spinors  $ \chi^{(0)}_{\uparrow} $ and $ \chi^{(0)}_{\downarrow} $ with the quantization axis in $ z $-direction of  the initial Cartesian system. Naturally, in such a case the expression (\ref{H_2D_0}) is not diagonal with respect to spin  $ \sigma = \uparrow ,\downarrow $. The transformation from the operators  $ a_{n,\mathbf{k}_{\perp},\sigma}^{\dagger} $ and $ a_{n,\mathbf{k}_{\perp},\sigma} $ to the operators $ a_{\mathbf{k}_{\perp},\uparrow} $ and $ a_{\mathbf{k}_{\perp},\downarrow} $ is performed by the unitary matrix
\[
\left( \begin{array}{c}
a_{\mathbf{k}_{\perp},+} \\
a_{\mathbf{k}_{\perp},-}
\end{array} \right) =
 \left(  \begin{array}{c} 
\cos \frac{\vartheta_{\mathbf{k}_{\perp}}}{2}  \quad \quad e^{-i\varphi_{\mathbf{k}_{\perp}}} \sin \frac{\vartheta_{\mathbf{k}_{\perp}}}{2}  \\
-e^{i\varphi_{\mathbf{k}_{\perp}}} \sin \frac{\vartheta_{\mathbf{k}_{\perp}}}{2} \quad  \cos \frac{\vartheta_{\mathbf{k}_{\perp}}}{2}   
\end{array} \right)
\left( \begin{array}{c}
a_{\mathbf{k}_{\perp},\uparrow} \\
a_{\mathbf{k}_{\perp},\downarrow}
\end{array} \right) ,
\]
where the relations are used 
\[
\sum_{\sigma} \chi^{(0)}_{\mathbf{k}_{\perp},\sigma} a_{\mathbf{k}_{\perp},\sigma} = \left(  \begin{array}{c} 
\cos \frac{\vartheta_{\mathbf{k}_{\perp}}}{2} a_{\mathbf{k}_{\perp},+} - e^{-i\varphi_{\mathbf{k}_{\perp}}} \sin \frac{\vartheta_{\mathbf{k}_{\perp}}}{2} a_{\mathbf{k}_{\perp},-} \\
e^{i\varphi_{\mathbf{k}_{\perp}}} \sin \frac{\vartheta_{\mathbf{k}_{\perp}}}{2} a_{\mathbf{k}_{\perp},+} + \cos \frac{\vartheta_{\mathbf{k}_{\perp}}}{2}  a_{\mathbf{k}_{\perp},-} 
\end{array} \right) = \left( \begin{array}{c}
a_{\mathbf{k}_{\perp},\uparrow} \\
a_{\mathbf{k}_{\perp},\downarrow}
\end{array} \right) .
\]
With account of the explicit form (\ref{F(k)}) of functions $ f \left( \mathbf{k}_{\perp} \right) $ in the representation of the operators $ a_{\mathbf{k}_{\perp},\sigma}^{\dagger} $ and $ a_{\mathbf{k}_{\perp},\sigma} $ ($ \sigma = \uparrow ,\downarrow $), the Hamiltonian (\ref{H_2D_0}) takes the form  
\begin{equation}
\label{H_2D_1} 
\mathrm{H}_{2D}^{(0)} = \sum_{\mathbf{k}_{\perp}} \left( a_{\mathbf{k}_{\perp},\uparrow}^{\dagger} \, a_{\mathbf{k}_{\perp},\downarrow}^{\dagger} \right) \left( E \left( 0 \right) + \frac{\hbar^{2} \mathbf{k}_{\perp}^{2} }{2m} + \hat{V}_{SO}^{(2D)} \right) \left( \begin{array}{c}
a_{\mathbf{k}_{\perp},\uparrow} \\
a_{\mathbf{k}_{\perp},\downarrow} ,
\end{array} \right)  
\end{equation}
which contains SOI determined by the confinement potential  
\begin{equation}
\label{H_SO2D} 
\hat{V}_{SO}^{(2D)} = - \lambda_{SO} \frac{ \left( \left[ \mathbf{e}_{z} \times \mathbf{k}_{\perp} \right] \mathbf{r}^{(0)}_{ \mathbf{k}_{\perp} } \right) }{1 + \mathbf{e}_{z} \mathbf{r}^{(0)}_{ \mathbf{k}_{\perp} } } \mathbf{r}^{(0)}_{ \mathbf{k}_{\perp}} \bm{\hat{\sigma}} .
\end{equation}
The latter expression evidently shows the mentioned above dependence of SOI on the wave vector components and angles $ \theta_{\mathbf{k}_{\perp}}  $,  $ \varphi_{\mathbf{k}_{\perp}} $, i.e. on the spatial and spin degrees of freedom, respectively.  

Expression (\ref{H_SO2D}) indicates directly that SOI is absent if the quantization axis is parallel to one of the vectors $ \mathbf{e}_{z} $ or  $ \mathbf{k}_{\perp} $. As it follows from Eq. (\ref{r_mu,r_S_2D}), in the first case the equality $ \mathbf{r}^{(0)}_{ \mathbf{k}_{\perp}} = \mathbf{r}_{\bm{\mu}} = \mathbf{e}_{z} $ is valid, and the invariant (\ref{Inv_g2}) reduces to $ z $-component of invariant (\ref{mu}): $ \hat{\mathcal{I}}_{gen} = \hat{\mu}_{z} $. In the second case at  $ \mathbf{r}^{(0)}_{ \mathbf{k}_{\perp} } = \mathbf{r}_{\bm{\mathcal{S}}} \left( \mathbf{k}_{\perp} \right) = \mathbf{k}_{\perp} /\vert \mathbf{k}_{\perp} \vert $, this invariant becomes  $ \hat{\mathcal{I}}_{gen} = \mathbf{k}_{\perp} \bm{\hat{\mathcal{S}}} $. 

Above the vector $ \mathbf{r}^{(0)}_{ \mathbf{k}_{\perp}} $  has been expressed in the Cartesian coordinate system in the form of the expansion with respect to unit vectors $ \mathbf{e}_{x} $, $ \mathbf{e}_{y} $ and $ \mathbf{e}_{z} $. In a 2D system it is convenient to use a local reper with another three unit vectors 
\begin{equation}
\label{basis_k} 
\mathbf{e}_{1} = \mathbf{k}_{\perp} /\vert \mathbf{k}_{\perp} \vert , \quad \mathbf{e}_{2} = \mathbf{e}_{z} \times \mathbf{e}_{1} , \quad \mathbf{e}_{3} = \mathbf{e}_{z} ,
\end{equation}
where $ \mathbf{r}^{(0)} _{ \mathbf{k}_{\perp}} = \gamma_{1}\left( \mathbf{k}_{\perp} \right) \mathbf{e}_{1} + \gamma_{2}\left( \mathbf{k}_{\perp} \right) \mathbf{e}_{2} + \gamma_{3}\left( \mathbf{k}_{\perp} \right) \mathbf{e}_{3} $ and the coefficients  $ \gamma_{j} $ ($ j=1,2,3 $), playing the role of the guiding cosines. It follows from expression  (\ref{H_SO2D}) that SOI attains its maximum value at $ \mathbf{r}^{(0)}_{ \mathbf{k}_{\perp} } = \mathbf{e}_{2} $, where $ \mathbf{e}_{2} $ is defined by (\ref{basis_k}). This corresponds to the invariant  $ \hat{\mathcal{I}}_{gen} = \hat{\epsilon}_{z} $. On the contrary, SOI vanishes,  $ \hat{V}_{SO}^{(2D)} = 0 $, if $ \mathbf{r}^{(0)}_{ \mathbf{k}_{\perp} } \perp \mathbf{e}_{2} $. This convincingly demonstrates the possibility of continuous changes of the SOI value under the smooth changes of the spin state, when, according to (\ref{Inv_g2}), the generalized invariant is represented in the form of a linear combination of operators $ \bm{\hat{\mu}} $ and $ \bm{\hat{\mathcal{S}}} $. This combination has to be either predefined or given for each concrete situation, and in the general case this linear combination has fundamental meaning, only.  

%\section{ Conclusions} \label{Concl}
\section{7. Conclusions}

This study demonstrates that SOI conventionally used in SchE, does not describe all possible spin electron  states. If the potential operator commutes with one of the spin invariants, the SOI operator has to be generalized to take into account all possible states. This is reflected in the fact that the vector $ \mathbf{r}^{(0)}_{ \mathbf{k}_{\perp}} $ is not fixed \textit{a priori}, and, hence, spin variables in the operator (\ref{H_2D_1}) can take \textit{arbitrary} values. 

Our results prove the spin lability of 2D electrons and show that their spectrum depends on the direction of the quantization axis which is determined by such factors as carriers concentration, form of the potential, presence of external fields, etc.  In a free 2D electron gas the most advantageous energy state is the state corresponding to the invariant $ \hat{\epsilon}_{z} $, when SOI operator reduces to Rashba SOI, as it has been indicated in \cite{AoP2}.  This interaction arises when only zero-harmonic of the potential (\ref{V_Four}) is taken into account, and, in this sense it can be considered as the zero approximation of SOI in real 2D (or quasi-3D) systems.  The Hamiltonians (\ref{H_2D_0}), (\ref{H_2D_1}) describe free 2D electrons, whose two-dimensional behavior in a homogeneous isotropic plane is determined by the QW, which means their localization in the potential  $ V_{0}\left( z \right) $. The full Hamiltonian includes also the periodical part,  $ V_{\perp}(\mathbf{r}_{\perp},z) $, and is given by the expression  $ \mathrm{H} \leftarrow  \mathrm{H}_{2D} = \mathrm{H}_{2D}^{(0)} + \mathrm{V}_{per} $, where the expression for $\mathrm{V}_{per} $ is written in (\ref{H_2D_tot}).   

After the transition to the operators  $ a_{n,\mathbf{k}_{\perp},\sigma}^{\dagger} $ and $ a_{n,\mathbf{k}_{\perp},\sigma} $ with the help of the unitary transformation (\ref{transf1_2D}), the Hamiltonian  $ \mathrm{H}_{2D} $ describes 2D electrons in a periodic field with account of SOI in the form (\ref{H_SO2D}), and from the potential  $ \mathrm{V}_{per} $ of crystal lattice. In literature, operator $ \hat{V}_{SO}^{(2D)} $ is usually called Rashba SOI, and operator which results from the potential $ \mathrm{V}_{per} $, is called Dresselhaus SOI.  

The explicit calculation of SOI with account of periodic  potential requires special consideration and is not the subject of the present paper. Nevertheless, it is worth mentioning that calculation of quasi-particle bands in such potential even without account of SOI is a rather difficult problem which is based on a spatial symmetry group of the crystal and can be done only approximately. For bulk crystals when  $ V_{0} = \mathrm{const} $ and SOI is not present in the zero approximation, spin-orbit splitting of such bands has been analyzed in details by Dresselhaus in Ref. \cite{Dresselhaus} for  zinc blender crystals and by Rashba and Sheka in Ref. \cite{Rashba1959} for  wurtzite-type crystals. These papers were based on the group theory using the   $ \mathbf{k} \mathbf{p} $ perturbation approximation with account of point symmetries of the Brillouin zone. The relativistic corrections to the spinors were taken into account in the first order perturbation theory. Such calculations are rather cumbersome and  accurate calculation of the generalized SOI based on the secondary quantization representation taking into account periodic potential, will be reported elsewhere.

\vspace{0.5cm}
\textbf{Acknowledgements}. We express our sincere thanks to E.I Rashba for reading the manuscript and  useful comments. This work has been done under the Fundamental Research Projects No 0117U000236 and No 0117U000240 of the Department of Physics and Astronomy of the  National Academy of Sciences of Ukraine.

\end{document}